\documentclass[preprint,onecolumn,nofootinbib]{revtex4}
\pdfoutput=1
\usepackage{graphicx}
\usepackage{amsmath}
\usepackage{amsfonts}
\usepackage{amssymb,ulem}
\usepackage{color}%
\usepackage{dcolumn}

\usepackage{MnSymbol,wasysym}
\usepackage{braket}

\usepackage{eurosym}
\usepackage{calrsfs}
\usepackage[usenames,dvipsnames,svgnames]{xcolor}

\newcommand{\RNum}[1]{\uppercase\expandafter{\romannumeral #1\relax}}
\usepackage[colorlinks=true,linkcolor=blue,urlcolor=blue,filecolor=black,citecolor=red,pdfstartview=FitV,pdftitle={},pdfsubject={},pdfkeywords={},pdfpagemode=None,bookmarksopen=true]{hyperref}

\begin{document}
\baselineskip=0.5 cm
\title{Evolutions of entanglement and complexity  after a thermal quench in massive gravity theory }
\author{Yu-Ting Zhou$ ^{1,2}$}
\email{constaantine@163.com}
\author{Mahdis Ghodrati$ ^{2,3}$}
\email{mahdisg@yzu.edu.cn}
\author{Xiao-Mei Kuang$ ^{2,3}$}
\email{xmeikuang@yzu.edu.cn}
\author{Jian-Pin Wu$ ^{2,3}$}
\email{jianpinwu@yzu.edu.cn}
\affiliation{$^1$ College of Mathematics and Science, Yangzhou University, Yangzhou 225009, China}
\affiliation{$^2$ Center for Gravitation and Cosmology, College of Physical Science and Technology,
Yangzhou University, Yangzhou 225009, China}
\affiliation{$^3$ School of Aeronautics and Astronautics, Shanghai Jiao Tong University, Shanghai 200240, China}
\date{\today }
\begin{abstract}
\vspace*{0.6cm}
\baselineskip=0.5 cm

We study the evolution of the holographic entanglement entropy (HEE) and the holographic complexity (HC) after a thermal quench in $1+1$ dimensional boundary CFTs dual to massive BTZ black holes. The study indicates how the  graviton mass $m_g$, the charge $q$, and also the size of the boundary region $l$  determine the evolution of the HEE and HC. We find that for small $q$ and $l$, the evolutions of the HEE and the HC is a continuous function. When $q$ or $l$ is tuned larger, the discontinuity emerges, which could not observed in the neutral AdS$_3$ backgrounds. We show that, the emergence of this discontinuity is a universal behavior in the charged massive BTZ theory.
With the increase of graviton mass, on the other hand, no emergence of the discontinuity behavior for any small $q$ and $l$ could be observed.
We also show that the evolution of the HEE and HC both become stable at later times, and $m_g$  speeds up reaching to the stability during the evolution of the system. Moreover, we show that $m_g$ decreases the final stable value of HEE but raises the  stable value of HC.
Additionally, contrary to the usual picture in the literature that the evolution of HC has only one peak, for big enough widths,  we show that graviton mass could introduce two peaks during the evolution. However, for large enough charges the one peak behavior will be recovered again.
We also examine the evolutions of HEE and HC growths at the early stage, which an almost linear behavior has been detected.

\end{abstract}

\maketitle

\tableofcontents

\section{Introduction}

The interdisciplinary connections of quantum information, condensed matter and quantum gravity, and the many possible applications are becoming increasingly more important in the realm of the theoretical physics and quantum information in this era. Specifically, holography \cite{Maldacena:1997re,Gubser:1998bc, Witten:1998qj} has been shown to play a remarkable and crucial role in this integration. Holography  helps to calculate many physical quantities in various CFTs and specifically in strongly correlated systems.

Specifically, two main quantities of entanglement entropy (EE) and complexity are very important concepts in the information theory both of which  can be calculated using holography, although it is extremely difficult to compute them in the field theory sides when the degrees of freedom of the system become large. Fortunately though, using holography for these two important quantities and recently for some related quantities, some elegant and simple geometric descriptions from gravity side have been provided.  For instance, 
one of the most important quantum information quantity for mixed states is entanglement of purification (EoP) introduced in \cite{Tarhel:0202044}, where its holographic dual has been considered to be the minimal entanglement wedge cross section \cite{Takayanagi:2017knl}. Additionally, bit thread formalism for studying EoP has been addressed in \cite{Bao:2019wcf,Harper:2019lff,Ghodrati:2019hnn,Du:2019emy}.  Another one is complexity of purification(CoP)  introduced in \cite{Agon:2018zso, Ghodrati:2019hnn}, which  is the minimum number of gates needed to purify a mixed state. Next but not the last interesting quantum information quantity is the logarithmic negativity, which is a quantum entanglement measure for mixed quantum states and only captures the ``quantum correlations" with the nature of ``entanglement"\cite{Plenio2005}. Its holographic dual has been studied in \cite{Kudler-Flam:2018qjo, Kusuki:2019zsp}.

In particular, it has been proposed that in holographic framework, the EE for a subregion on the dual boundary is proportional to the minimal surface in the bulk geometry, for which is being called the Hubeny Rangamani-Takayanagi (HRT) surface \cite{Takayanagi:2012kg,Hubeny:2007xt}. One of the most important applications of holographic entanglement entropy (HEE) and the studies of HRT surfaces is to diagnose and study various holographic phase transitions, for example see \cite{Ling:2015dma,Ling:2016wyr,Ling:2016dck,Pakman:2008ui,Kuang:2014kha,Klebanov:2007ws,Zhang:2016rcm,
Zeng:2016fsb,Guo:2019vni}.

In addition, quantum complexity measures how many quantum gates are required to prepare, up to a specific precision,  the target state from the initial state in any quantum circuit model \cite{Watrous:2009,Osborne:2012,Gharibian:2014}. However, its exact definition in the quantum fields theory is notoriously difficult and the complete definition is unclear yet. Some recent progress, though, in this regard have been made in \cite{Chapman:2017rqy,Jefferson:2017sdb,Khan:2018rzm,Hackl:2018ptj,Guo:2018kzl}.
The difficulty mainly arrises due to the fact that, the Hilbert space is so large, the degrees of freedom of the system are infinite, and there exist some ambiguities on how to define the unitary operations and also the reference state. In fact, instead of discrete gates, a continuos definition is called for. On the other hand, holography could recently provide an alternative, well-defined method to study the computational complexity of different dual field theories.

In the holographic framework, there are two different proposals to evaluate the computational complexity. One is the CV conjecture (Complexity=Volume) \cite{Stanford:2014jda,Susskind:2014jwa},
and the other is being called the CA conjecture (Complexity=Action) \cite{Brown:2015bva,Brown:2015lvg}.
The CV conjecture proposes that the holographic complexity (HC) is proportional to the volume of
a codimension-one hypersurface with the AdS boundary and the HRT surface. While to use the CA conjecture, one should identify the HC with the gravitational action evaluated on the Wheeler-DeWitt patch in the bulk. In this paper, we shall follow the CV conjecture and study its evolution under a thermal quench.

The fascinating point is that the study of HEE and HC, could provides us with more power and tools to explore the nature of the spacetime, in particular the physics of the black hole horizon. Specifically, while studying the information paradox in black holes, the authors of \cite{Susskind:2014rva, Susskind:2014moa} have found that the entanglement entropy could not be enough to understand the black hole horizon Therefore, they proposed the ER=EPR conjecture and argued that the creation of the firewall behind the horizon is essentially a problem of ``quantum computational complexity" \cite{Stanford:2014jda}. This was an example for how studying the evolution of information quantities such as HEE and HC could provide us more with information about the nature of the black hole horizon and even its thermal and entanglement structures.

Therefore, the evolution of the HEE and HC has been explored in various dynamical backgrounds such as Vaidya-AdS spacetime \cite{Chen:2018mcc} and in Einstein-Born-Infeld theory \cite{Ling:2018xpc}. The authors of those works investigated the HEE and HC under a thermal quench in the related gravitational background. This kind of quench process in the dual boundary field theory is described holographically by the black hole formation from the gravitational collapse, and it is widely employed as an effective model to study thermalization process, see for example \cite{Balasubramanian:2010ce,Balasubramanian:2011ur} as a review. The study on the evolution of subregion complexity has been generalized to chaotic system \cite{Yang:2019vgl} and dS boundary \cite{Zhang:2019vgl}.  The evolutions of HEE and HC for quantum quench have also been studied in \cite{Leichenauer:2015xra,Leichenauer:2016rxw,Moosa:2017yiz} and therein.

In this paper, we shall investigate the HEE and HC under a thermal quench in three dimensional massive gravity theory. We shall separately explore the effects from the mass of graviton and charge of black hole. Also, we study the joint, simultaneous effects from both of them.

Our paper is organized as follows. In section \ref{sec-II}, we introduce the general framework describing HEE and HC in Vaidya-AdS$_{3}$ spacetime. Then, in section \ref{sec-III}, we separately explore the effects from the mass of graviton and charge of black hole. Also, we study their  joint effects. Finally, in section \ref{sec-IV}, we summarize our results.

\section{Holographic setup of  HEE and HC in  Vaidya-AdS$_{3}$ theory}\label{sec-II}

In order to study the evolution of HEE and HC in $1+1$ dimensional field theory after a thermal quench via holography, we consider the Vaidya-AdS$_{3}$ spacetime with a planar horizion in terms of Poincare coordinate
\begin{eqnarray}\label{eq-metric0}
ds^2=\frac{1}{z^2}\left(-f(v,z)dv^2-2dzdv+dx^2\right)\,,
\end{eqnarray}
where $f(v,z)$ is the redshift function. Also, $v$ is the ingoing null trajectory, which coincides with the time coordinate $t$ on the conformal boundary. Note that in the limit $v\to -\infty$, the above metric reduces to a pure AdS$_3$ spacetime, while in the limit $v\to \infty$, it describes certain AdS BTZ black holes. These parameters will be explicitly fixed later.
\begin{figure}[ht!]
 \centering
  \includegraphics[width=7cm] {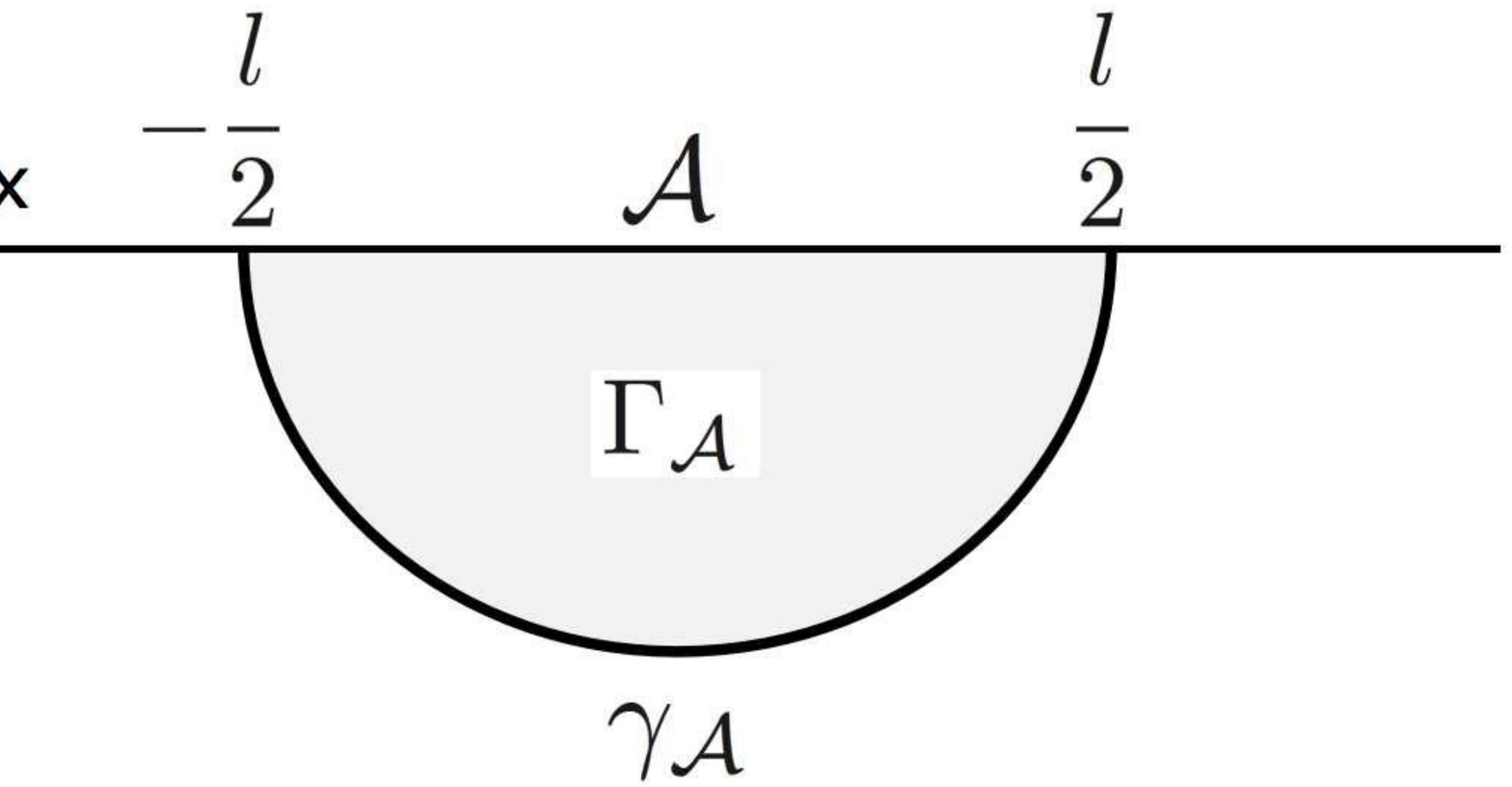}
  \caption{Geometrical description of the subregion $\mathcal{A}$ with width $l$. The $\gamma_{\mathcal{A}}$ denotes a codimension-one surface
  and the shadow region is the volume of hypersurface $\Gamma_{\mathcal{A}}$ with boundaries $\mathcal{A} $ and $\gamma_{\mathcal{A}}$.}
 \label{fig:strips}
\end{figure}

To study HEE and HC in the dynamical spacetime, we first consider the subregion as $\mathcal{A}=x \in\left(-\frac{l}{2}, \frac{l}{2}\right)$ with finite $l$ in an asymptotic AdS background. The setup is shown in Fig.\ref{fig:strips}.
It was proposed in \cite{Hubeny:2007xt} that in the dynamical spacetime, HEE  for a subregion $\mathcal{A}$ on the boundary could be captured by a codimension-two bulk surface with vanishing expansion of geodesics, i.e., the HRT surface $\gamma_{\mathcal{A}}$, while the corresponding HC
is proportional to the volume of a codimension-one hypersurface $\Gamma_{\mathcal{A}}$ with the boundaries  $\mathcal{A}$ and $\gamma_{\mathcal{A}}$.

So we will follow the strategy in \cite{Chen:2018mcc} to analytically derive the integral expressions of HEE via the minimal surface, and for calculating HC  we use the CV conjecture.
Due to the symmetry of the system, the corresponding extremal surface $\gamma_{\mathcal{A}}$ in the bulk can be parametrized as
\begin{eqnarray}
v=v(x), \quad z=z(x), \quad z(\pm l / 2)=\epsilon, \quad v(\pm l / 2)=t-\epsilon,
\end{eqnarray}
where $\epsilon$ is the cut-off. Then the induced metric on the surface is
\begin{eqnarray}
d s^{2}=\frac{1}{z^{2}}\left[-f(v, z) v^{\prime 2}-2 z^{\prime} v^{\prime}+1\right] d x^{2},
\end{eqnarray}
where the prime denotes the derivative with respect to $x$.
It is straightforward then to write down the area of the extremal surface as
\begin{eqnarray}
\operatorname{Area}\left(\gamma_{\mathcal{A}}\right)= \int_{-l / 2}^{l / 2} \frac{\sqrt{1-f(v, z) v^{\prime 2}-2 z^{\prime} v^{\prime}}}{z}d x.
\end{eqnarray}
To extract the HEE, one then has to minimize the above surface. The useful trick is to treat  the above  function as an action, and then the corresponding Lagrangian and Hamiltonian are
\begin{eqnarray}
\label{eq-lagrangian}\mathcal{L}_{S}&=&\frac{\sqrt{1-f(v, z) v^{\prime 2}-2 z^{\prime} v^{\prime}}}{z},\\
\label{eq-Hamitonion}\mathcal{H}_{S}&=&\frac{1}{z\sqrt{1-f(v, z) v^{\prime 2}-2 z^{\prime} v^{\prime}}}.
 \end{eqnarray}
Since  the Hamiltonian does not explicitly depend on the variable $x$, it is conserved. Beside, the symmetry of the surface gives us a turning point, $(z_{*}, v_{*})$, of the extremal surface  $\gamma_{\mathcal{A}}$ which is located at $x =0$.
So we could set
\begin{eqnarray}\label{eq-boundarycondition}
 v^{\prime}(0)=z^{\prime}(0)=0, \quad z(0)=z_{*}, \quad v(0)=v_{*}.
\end{eqnarray}
Subsequently, the conserved Hamitonian gives
 \begin{eqnarray}
1-f(v, z) v^{\prime 2}-2 z^{\prime} v^{\prime}=\frac{z_{*}^{2}}{z^{2}}.
 \end{eqnarray}
Combining the result from the derivative of the Lagrangian \eqref{eq-lagrangian} with respect to $x$, and the equations of motion for $z(x)$ and for  $v(x)$, we obtain a group of partial differential equations as
  \begin{eqnarray}
\label{eq-PDE01}0&=&-2+2 z v^{\prime \prime}+v^{\prime}\left[2 f(v, x) v^{\prime}+4 z^{\prime}-z v^{\prime} \partial_{z} f(v, z)\right],\\
\label{eq-PDE02}0&=& 2 f(v, z)^{2} v^{\prime 2}+f(v, z)\left[-2+4 v^{\prime} z^{\prime}-z v^{2} \partial_{z} f(v, z)\right] \nonumber\\ &&~-z\left[2 z^{\prime \prime}+v^{\prime}\left(2 z^{\prime} \partial_{z} f(v, z)+v^{\prime} \partial_{v} f(v, z)\right)\right].
   \end{eqnarray}

We have to solve the above equations using the boundary conditions \eqref{eq-boundarycondition} and extract the solutions of $v=\tilde{v}(x), z=\widetilde{z}(x)$ for the extremal surface $\gamma_{\mathcal{A}}$. The extremal surface $\gamma_{\mathcal{A}}$ is then simplified as
\begin{eqnarray}\label{eq-S0}
\operatorname{Area}\left(\gamma_{\mathcal{A}}\right)=2  \int_{0}^{l / 2} \frac{z_{*}}{\tilde{z}(x)^{2}} d x,
\end{eqnarray}
which gives the HEE of the subregion on the boundary. One should note that the surface does not live on a constant time slice for the general $f(v,z)$, as both $z_{*}$ and $\tilde{z}(x)$ will be changed by time.

We then derive the general expression of HC via CV conjecture in the background \eqref{eq-metric0} which has the same profile as HEE.
One should note that, the codimension-one extremal surface $\Gamma_{\mathcal{A}}$ is bounded by the surface $ \gamma_{\mathcal{A}}$ in the bulk.  In \cite{Chen:2018mcc}, It has been addressed that there are in fact two equivalent ways to describe $\Gamma_{\mathcal{A}}$ which is  parameterized by $v(z)$ or $z(v)$. Using the profile $z(v)$ is usually more convenient for the dynamical backgrounds, which we will consider in the following study, i.e., we parameterize the extremal bulk region $\Gamma_{\mathcal{A}}$ enclosed by  $v=\tilde{v}(x), z=\tilde{z}(x) $ via $ z=z(v)$.
Thus, the induced metric on $\Gamma_{\mathcal{A}}$ would be calculated as
 \begin{eqnarray}
 d s^{2}=\frac{1}{z^{2}}\left[-\left(f(v, z)+2 \frac{\partial z}{\partial v}\right) d v^{2}+d x^{2}\right],
 \end{eqnarray}
 and  the volume could be evaluated as
 \begin{eqnarray}
 V(\Gamma_{\mathcal{A}})=2  \int_{v_{*}}^{\tilde{v}(l / 2)} d v \int_{0}^{\tilde{x}(v)} \frac{d x}{z^2}\left[-f(v, z)-2 \frac{\partial z}{\partial v}\right]^{1 / 2},
 \end{eqnarray}
where $\tilde{x}(v)$ is the coordinate in the codimension-two extremal surface $\gamma_{\mathcal{A}}$.
Similarly, treating the above integral function as the Lagrangian, we obtain the equation of motion as
 \begin{eqnarray}\label{eq-eomForV}
 \begin{aligned} 0=&\left[4f(v, z)^{2}+8z^{\prime}(v)^{2}-3 z(v) z^{\prime}(v) \partial_{z} f(v, z)+f(v, z)\left(12 z^{\prime}(v)-z(v) \partial_{z} f(v, z)\right)\right.\\ &-z(v)\left(2 z^{\prime \prime}(v)+\partial_{v} f(v, z)\right) ] /\left[z(v)^{3}\left(-f(v, z)-2 z^{\prime}(v)\right)^{3 / 2}\right].
 \end{aligned}
  \end{eqnarray}

One may solve the above equation using the boundary condition which is determined by the codimension-two surface $\gamma_{\mathcal{A}}=(\tilde{v}(x), \tilde{z}(x))$ and $\mathcal{A}$. Alternatively, similar to the case in HEE, the solution to \eqref{eq-eomForV} could also be figured out by finding $\tilde{z}(\tilde{v})$ on the boundary $\gamma_{\mathcal{A}}$. Subsequently, the volume is rewritten as
\begin{eqnarray}\label{eq-V0}
V(\Gamma_{\mathcal{A}})=2  \int_{v_{*}}^{\tilde{v}(l / 2)} d v\left[-f(v, z(v))-2 \frac{\partial z}{\partial v}\right]^{1 / 2} z(v)^{-2} \tilde{x}(v),
\end{eqnarray}
which is dual to HC of the subregion in the boundary.

Once, for the strip subregion, we find the general expressions of HEE from equation \eqref{eq-S0}, and HC from equation\eqref{eq-V0}, both holographically and in the background of Vaidya-AdS$_{3}$ black hole, we could then go forward by numerically studying the ``evolutions" of HEE and HC for this specific geometry and setup.

\section{Evolution of HEE and HC in the massive charged BTZ black hole}\label{sec-III}

In this section, we study the evolution of HEE and HC after a thermal quench in the background of massive charged BTZ black hole.

\subsection{The general formulas of HEE and HC in massive charged BTZ black hole}

As for the theory, we choose the Einstein-Maxwell-massive gravity in three dimensional spacetimes, which is expressed in the following form \cite{Hendi:2016pvx}
\begin{equation}
\mathcal{I}=-\frac{1}{16\pi }\int d^{3}x\sqrt{-g}\left[ \mathcal{R}+2-F^{2}+m_{g}^{2}\sum_{i}^{4}c_{i}\mathcal{U}_{i}(g,h)\right]\,,
\label{Action}
\end{equation}%
where $c_i$ are constants and $m_g$ is the mass of the graviton. Also, $F=dA$ is the Maxwell field strength of gauge field $A$, and $h_{\mu\nu}$ is the reference metric, which is a symmetric tensor.
Plus, $\mathcal{U}_{i}$ are the polynomials of the eigenvalues of the matrix $\mathcal{K}_{\nu }^{\mu }=\sqrt{%
g^{\mu \alpha }h_{\alpha \nu }}$, and the forms are
\begin{eqnarray}\label{eq-Ui}
&&
\mathcal{U}_{1} =\left[ \mathcal{K}\right] ,\;\;\;\;\;\mathcal{U}_{2}=%
\left[ \mathcal{K}\right] ^{2}-\left[ \mathcal{K}^{2}\right] ,\;\;\;\;\;%
\mathcal{U}_{3}=\left[ \mathcal{K}\right] ^{3}-3\left[ \mathcal{K}\right] %
\left[ \mathcal{K}^{2}\right] +2\left[ \mathcal{K}^{3}\right] ,  \notag \\
&&\mathcal{U}_{4}=\left[ \mathcal{K}\right] ^{4}-6\left[ \mathcal{K}^{2}%
\right] \left[ \mathcal{K}\right] ^{2}+8\left[ \mathcal{K}^{3}\right] \left[
\mathcal{K}\right] +3\left[ \mathcal{K}^{2}\right] ^{2}-6\left[ \mathcal{K}%
^{4}\right] .
\end{eqnarray}
Similar to the case in \cite{Vegh:2013sk}, for the reference metric $h_{\mu\nu}$, we could choose the special case of $h_{\mu \nu }=diag(0,0,c^{2}h_{ij})$ and the corresponding polynomials $\mathcal{U}_{i}$ are evaluated as $\mathcal{U}_{1}=c/r$ and $\mathcal{U}_{2}=\mathcal{U}_{3}=\mathcal{U}_{4}=0$.
One should note that, the massive terms break the diffeomorphism symmetry of the bulk, which corresponds to momentum dissipation in the dual boundary field theory \cite{Vegh:2013sk,Blake:2013bqa}.

The action \eqref{Action} gives the following massive charged BTZ black hole geometry
\begin{equation}
ds^2=\frac{1}{z^2}[-f(z)dt^2+\frac{dz^2}{f(z)}+dx^ { 2 }]~~~\mathrm{with} ~~~f(z)=1-mz^{2}+m_{g}^{2}cc_{1}z+q^{2}z^{2}\ln z.
\label{eq-massiveBTZ}
\end{equation}%
Without loss of generality, we will set $c=c_1=1$ in the following study.
Then, we reformulate the above massive BTZ black hole metric into the Vaidya-AdS formula as
\begin{eqnarray}\label{eq-massiveBTZ-vaidya}
ds^2=\frac{1}{z^2}\left(-f(v,z)dv^2-2dzdv+dx^2\right),~~\mathrm{with}~f(v,z)=1-M(v)z^2+m_{g}^2z+Q(v)^{2}z^{2}\ln z\,.
\end{eqnarray}

We assume that the mass $M(v)$ and the charge $Q(v)$ of the black hole take the following form\cite{Balasubramanian:2011ur}
\begin{eqnarray}\label{eq-mv2}
M(v)&=&\frac{m}{2}(1+\tanh\frac{v}{v_0})\,,\nonumber\\
Q(v)&=&\frac{q}{2}(1+\tanh\frac{v}{v_0})\,,
\end{eqnarray}
where $v_0$ is the thickness of the shell.
It is then straightforward to check that in the limit of $v\to -\infty$, the background would describe a pure AdS space with corrections in graviton mass, and in the limit $v\to \infty$,
 the background reduces to the static solution \eqref{eq-massiveBTZ} in massive gravity.

Since the Hamiltonian \eqref{eq-Hamitonion} is conserved along the $x$ direction, we can obtain
 a relation between the length $l$ and $z_{*}$ as
\begin{eqnarray}
\int_{\epsilon}^{z_{*}}\left[\left(1-M(v)z^2+m_{g}^2z+Q(v)^{2}z^{2}\ln z\right)\left(\frac{z_{*}^{2 }}{z^{2 }}-1\right)\right]^{-1 / 2} d z=\int_{0}^{l / 2} d x=\frac{l}{2}\,.
\end{eqnarray}
Using the above relation, both the extremal surface and the subregion volume are explicitly derived as
\begin{eqnarray}
&&
\operatorname{Area}\left(\gamma_{\mathcal{A}}\right)=2 \int_{z_{*}}^{\epsilon} \frac{z_{*}}{z^{2 }}\left[\left(1-M(v)z^2+m_{g}^2z+Q(v)^{2}z^{2}\ln z\right)\left(\frac{z_{*}^{2 }}{z^{2 }}-1\right)\right]^{-1 / 2} d z\,,
\label{Area}
\\
&&
V\left(\Gamma_{\mathcal{A}}\right)=2 \int_{v}^{\tilde{v}(l / 2)} d v\left[-(1-M(v)z^2+m_{g}^2z+Q(v)^{2}z^{2}\ln z)-2 \frac{\partial z}{\partial v}\right]^{1 / 2} z(v)^{-2} \tilde{x}(v)\,.
\label{volume}
\end{eqnarray}

The group of differential equations \eqref{eq-PDE01} and \eqref{eq-PDE02} could also be explicitly expressed as

\begin{eqnarray}
0&=& \frac{1}{4} v'(x)^2 \left(-4 m_{g}^2 z(x)+q^2 z(x)^2 \left(\tanh \left[\frac{v(x)}{v_{0}}\right]+1\right)^2-8\right) \nonumber\\ && -2 z(x) v''(x) -4 v'(x) z'(x)+2 \\
0&=&\frac{1}{16}(2(4+4m_{g}^{2}z(x)+(\tanh \left[\frac{v(x)}{\text{v0}}\right]+1))(-2m+q^{2}\ln[z(x)]+q^{2}\ln[z(x)]\tanh\left[\frac{v(x)}{v_{0}}\right])z(x)^{2})^{2}  \nonumber\\ &&
v'(x)^{2}-(4+4m_{g}^{2}z(x)+(1+\tanh\left[\frac{v(x)}{v_{0}}\right])(-2m+q^{2}\ln[z(x)]+q^{2}\ln[z(x)]\tanh\left[\frac{v(x)}{v_{0}}\right])z(x)^{2})\nonumber\\ &&
 (8+4m_{g}^{2}z(x)v'(x)^{2}+(1+\tanh\left[\frac{v(x)}{v_{0}}\right])\nonumber\\ &&
(-4m+q^{2}+2q^{2}\ln\left[z(x)\right]+q^{2}(1+2\ln\left[z(z)\right])\tanh\left[\frac{v(x)}{v_{0}}\right])z(x)^{2}v'(x)^{2} -16v'(x)z'(x))+\nonumber\\ &&
\frac{1}{v_{0}}8z(x)(\text{sech}\left[\frac{v(x)}{v_{0}}\right]^{2}(m-q^{2}\ln\left[z(x)\right]-q^{2}\ln\left[z(x)\right]\tanh\left[\frac{v(x)}{v_{0}}\right])z(x)^{2}v'(x)^{2}-\nonumber\\ &&
v_{0}(1+\tanh\left[\frac{v(x)}{v_{0}}\right])(-4m+q^{2}+2q^{2}\ln\left[z(x)\right]+q^{2}(1+2\ln\left[z(x)\right])\tanh\left[\frac{v(x)}{v_{0}}\right])-\nonumber\\ &&
v_{0}(1+\tanh\left[\frac{v(x)}{v_{0}}\right])(-4m+q^{2}+2q^{2}\ln\left[z(x)\right]+q^{2}(1+2\ln\left[z(x)\right])\tanh\left[\frac{v(x)}{v_{0}}\right])z(x)v'(x)z'(x)-\nonumber\\ &&
4v_{0}(m_{g}^{2}v'(x)z'(x)+z''(x)))).
\end{eqnarray}
We can numerically solve the above equations with the following boundary conditions,
\begin{eqnarray}\label{eq-bdys}
v^{\prime}(0)=z^{\prime}(0)=0\,,\,\, z(0)=z_{*}\,,\,\, v(0)=v_{*}\,,\,\,  z(l / 2)=\epsilon\,,\,\,  v(l / 2)=t-\epsilon\,.
\end{eqnarray}

Once the solution to the above equations is at hands, we can read off the HEE and HC from Eq.\eqref{Area} and Eq.\eqref{volume}, respectively. Note that both HEE and HC are divergent.
However, here we are only interested in the change of the HEE or the HC during the quench.
Therefore, we could define some finite quantities for HEE and HC by subtracting the vacuum part which is dual to the AdS geometry. Then, we  define the following finite and well-defined quantities

\begin{eqnarray}
&&
S=\frac{\operatorname{Area}(\gamma_{\mathcal{A}})-\operatorname{Area}_{\operatorname{AdS}}(\gamma_{\mathcal{A}})}{2l}
\,,\label{subtractHEE}
\\
&&
C=\frac{\operatorname{V}(\Gamma_{\mathcal{A}})-\operatorname{V}_{\operatorname{AdS}}(\Gamma_{\mathcal{A}})}{2l}\,.
\label{subtractHC}
\end{eqnarray}

Next, we  present the numerical results for the evolutions of these two quantities after quench.
We first focus on the neutral case, i.e., $q=0$, and we study the effects from the massive term in subsection \ref{sub-neutral}.
Later, we explore the effects of the charge $q$ on the evolutions of HEE and HC by turning off the mass term in subsection \ref{sub-charge}.
Finally, the joint effects of the charge of black hole and massive graviton will be presented in subsection \ref{sub-massive-charge}.

\subsection{HEE and HC in the neutral massive BTZ black hole}\label{sub-neutral}

\begin{figure}[ht!]
 \centering
  \includegraphics[width=7.8cm] {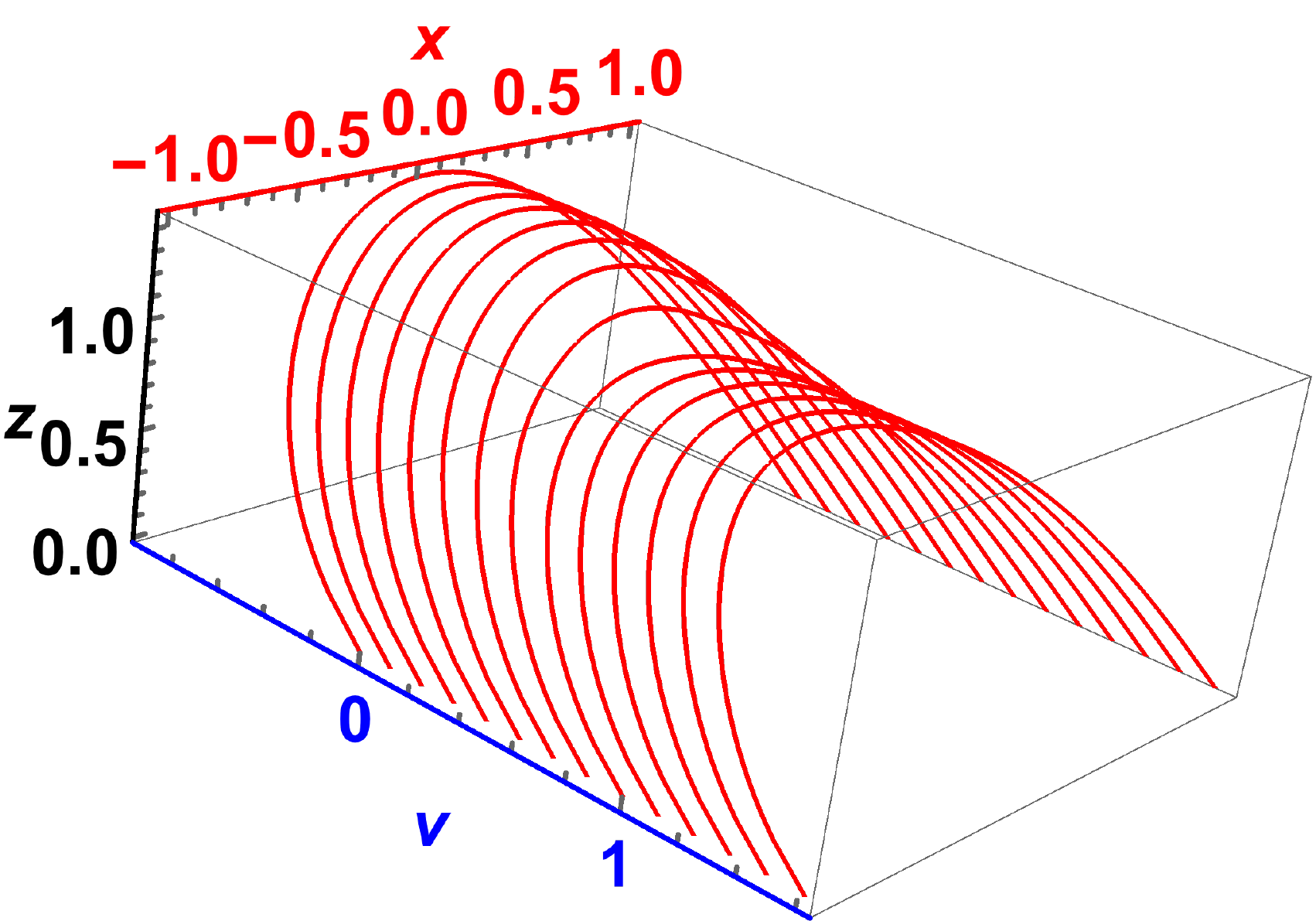}\ \hspace{0.5cm}
   \includegraphics[width=7.8cm] {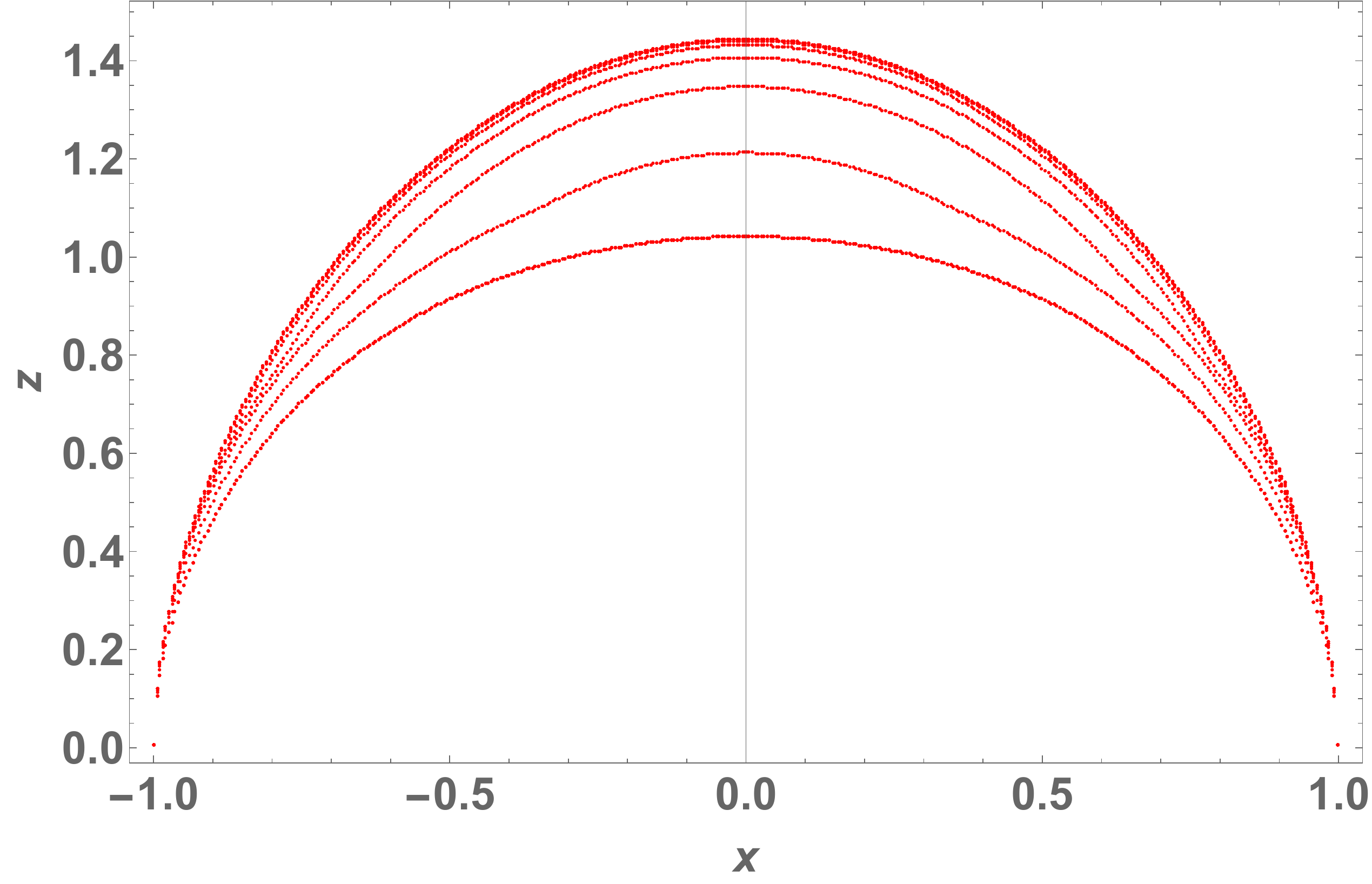}\ \\
  \caption{The evolution of extremal surface $\gamma_{\mathcal{A}} = (\tilde{z}(x), \tilde{v}(x))$ in the background of the neutral massive BTZ black hole.
  Here, we have set $m=1$, $q=0$, $m_g=1$, $v_0=0.01$ and $l = 2$.
  The left plot shows the evolution in $(x, v, z)$ space and the right plot shows the corresponding projection in $(x,z)$ plan.
  The evolution is from left to right in the left plot and from top to bottom in the right plot.}
 \label{fig:3dzvxSEEandzxSEE}
\end{figure}
\begin{figure}[ht!]
 \centering
  \includegraphics[width=5cm]{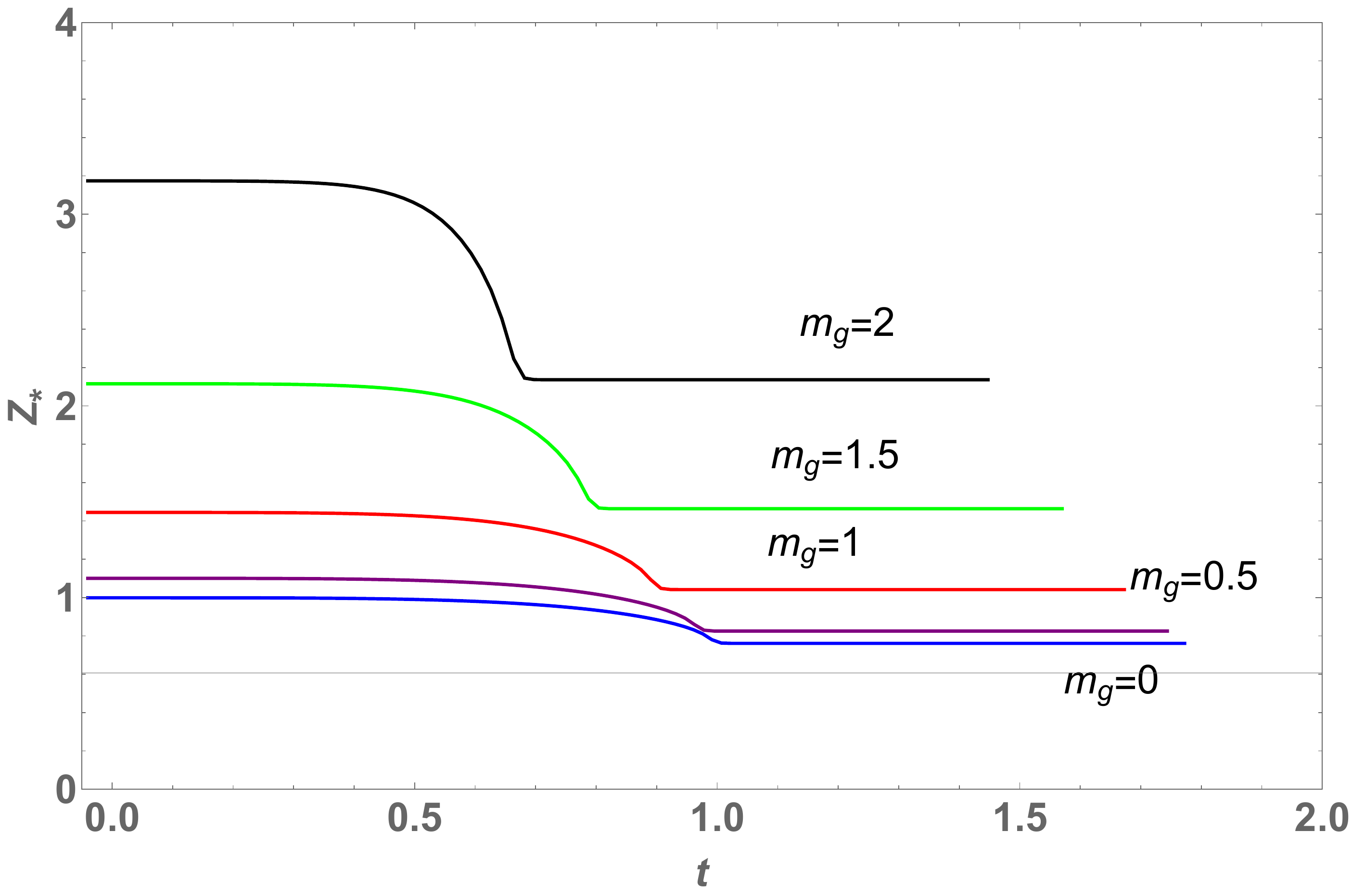}\ \hspace{0.1cm}
  \includegraphics[width=5cm]{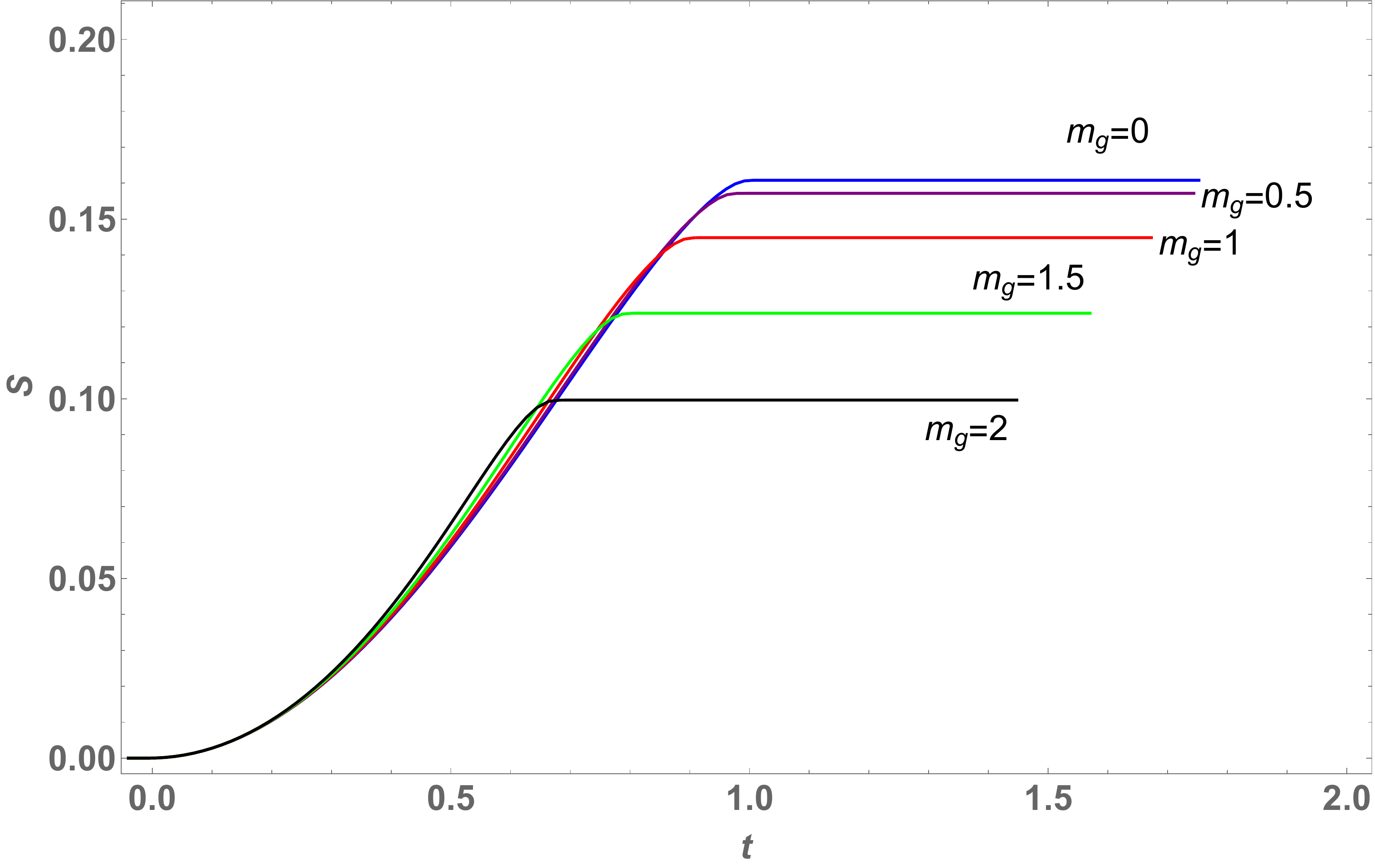}\ \hspace{0.1cm}
  \includegraphics[width=5cm]{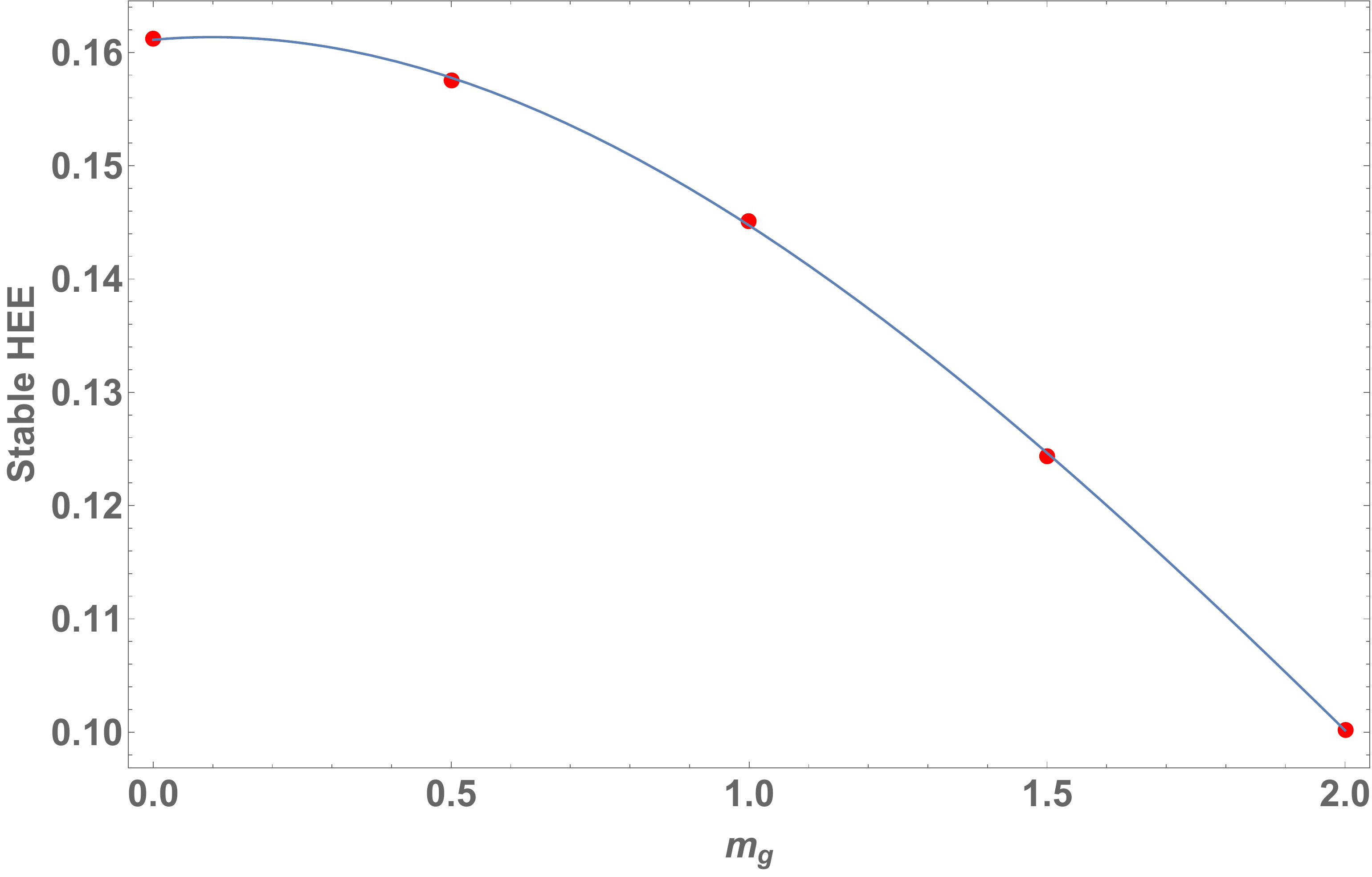}
	  \caption{Left plot: The evolution of $z_{*}$ for different $m_g$.
	  Middle plot: The evolution of the HEE for different $m_{g}$. Right plot: The relation between the stable value of the HEE at late time and $m_g$.
	  Here we have set $l$=2, $q=0$, $m=1$ and $v_{0}=0.01$.}
 \label{fig:SEEvsT}
\end{figure}

In this subsection, we mainly explore the effect from the massive graviton and so we turn off the charge of black hole, i.e., set $q=0$.
 We also fix $m=1$ and $v_{0}=0.01$. First, the effect of graviton mass for small region, i.e, $l=2$, and then for larger regions, $l=10$ are studied.

To gain an intuitive understanding of the evolution of HEE, we first explore the evolution of the HRT surface $\gamma_{\mathcal{A}}$. The left plot in Fig.\ref{fig:3dzvxSEEandzxSEE} exhibits the evolution in $(x,v,z)$ space, in which $\gamma_{\mathcal{A}}$ evolves from left to right. In the right plot of Fig.\ref{fig:3dzvxSEEandzxSEE}, the corresponding projection in $(x,z)$ plan is shown, in which the evolution is from top to bottom.
It is obvious that $\gamma_{\mathcal{A}}$ evolves smoothly from the initial state to the final state,
which is similar with the case in the Einstein gravity in \cite{Chen:2018mcc}.

Quantitatively, we also show the evolution of the turning point $z_*$ of the HRT for different $m_g$ in the left plot in Fig.\ref{fig:SEEvsT}.
One could see that, for the fixed value of $m_g$, the turning point $z_{*}$ is almost a constant function at the early stages of its evolution. This constant is different for different values of $m_g$.
This is because at the limit of $t\to 0$, i.e., $v\to -\infty$, the AdS background geometry is corrected by the mass parameter of graviton.
After such early stage, then $z_{*}$  rapidly decreases and finally enters into a stable stage, which is again  a constant.
In particular, at the limit of $t\to 0$ (i.e., $v\to-\infty$), one can see that the HEE vanishes.  The reason is just because the HEE dual to the AdS geometry has been subtracted  (see Eq.\eqref{subtractHEE}).
As the time evolves, the HEE climbs up monotonically and finally reaches to a stable value, which depends on the value of $m_g$.

There are two main characteristics for the $z_{*}$ and the HEE for different values of $m_g$.
These behaviors are summarized as follows:
\begin{itemize}
	\item With larger values for the graviton mass $m_g$, the two quantities of $z_{*}$ and HEE both reach the stability faster than with the small values of $m_g$. So as the graviton mass becomes larger, the corresponding boundary theory saturates into equilibrium faster. This result is similar to the effects of $m_g$ on the geodesic probe during the thermalization process,  studied in \cite{Hu:2016mym}. Those results show that the inhomogeneity of the boundary field theory  introduced by using $m_g$ in the bulk, makes the thermalization to occur faster, which here shown to be true for complexity as well.

	\item The stable value of $z_{*}$ and HEE, in the final stage and at the later times, are different for various values of $m_g$. As $m_g$ increases, the stable value of the turning point, $z_*$, increases, while HEE decreases. We quantitatively exhibit the relation between the stable value of the HEE and $m_g$ in Fig.\ref{fig:SEEvsT}.
\end{itemize}

\begin{figure}[ht!]
 \centering
  \includegraphics[width=8.5cm] {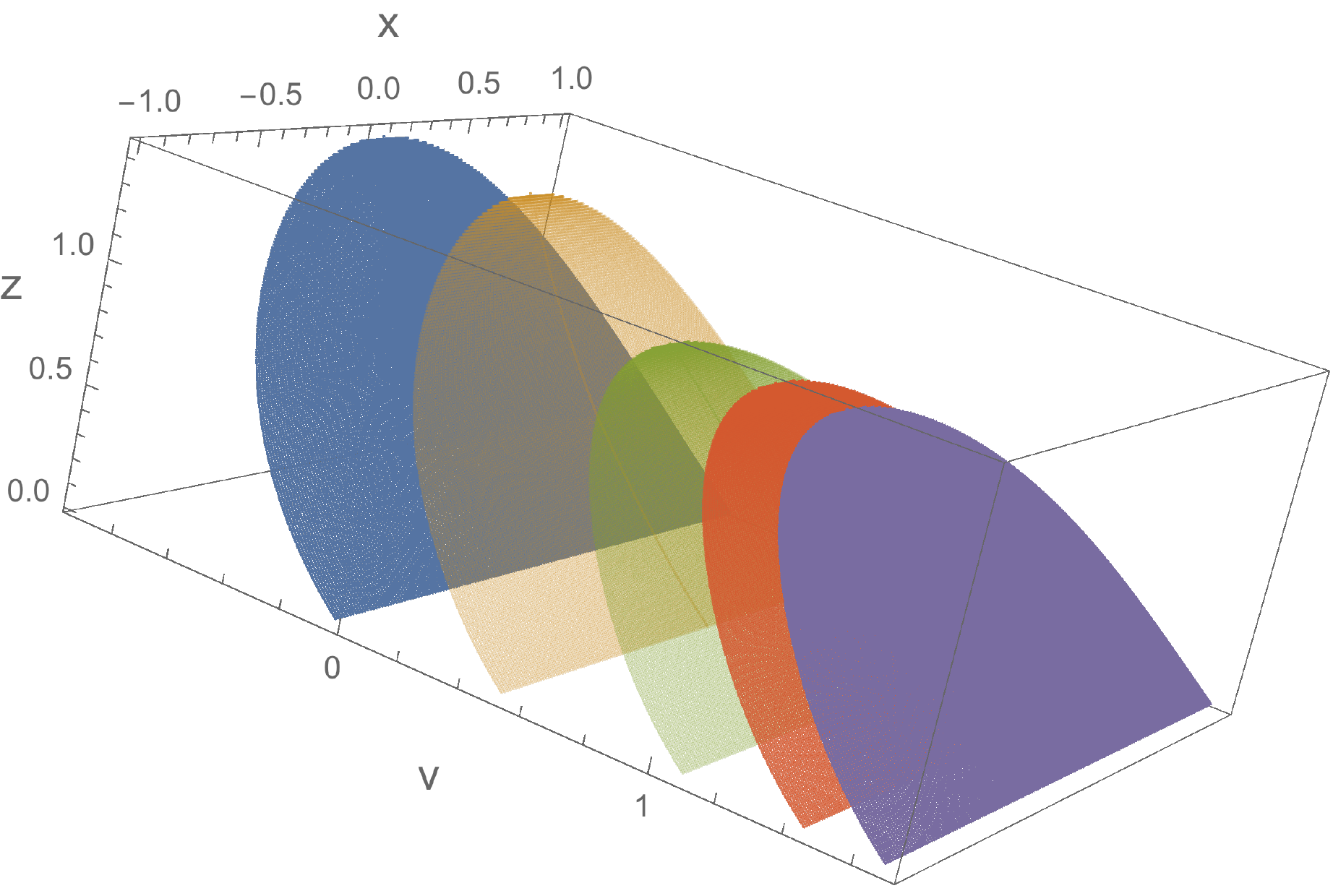}
  \caption{The evolution of the codimension-one extremal surface $V(\Gamma_{\mathcal{A}})$
  which characterizes the subregion complexity bounded by the HRT surface $\gamma_{\mathcal{A}}$ and the boundary area $\mathcal{A}$.
  Here, we have set $m=1$, $q=0$, $m_g=1$, $v_0=0.01$ and $l = 2$.}
 \label{fig:3dzvxC}
\end{figure}

\begin{figure}[ht!]
 \centering
  \includegraphics[width=5cm] {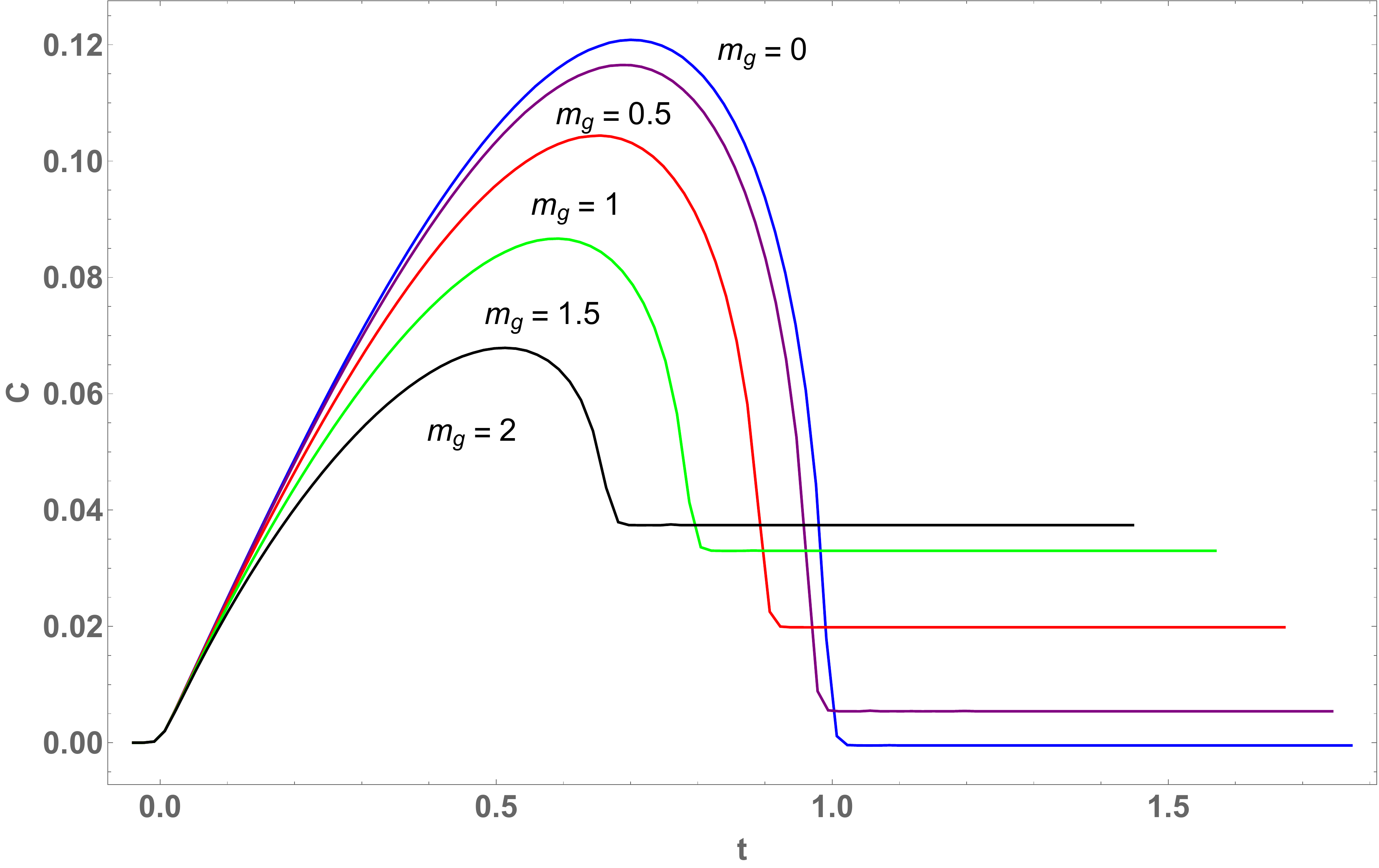}\ \hspace{0.1cm}
    \includegraphics[width=5cm] {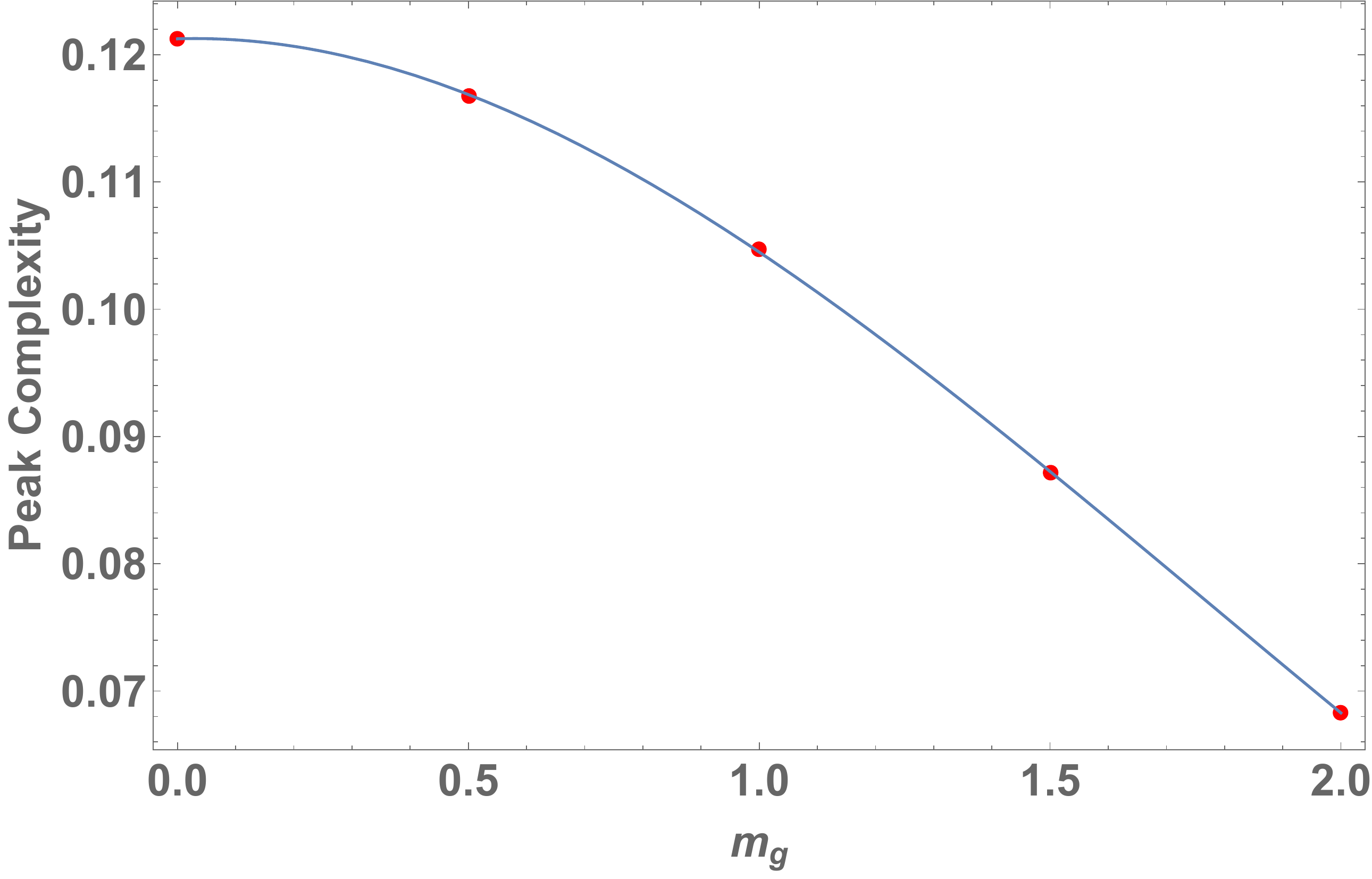}\ \hspace{0.1cm}
   \includegraphics[width=5cm] {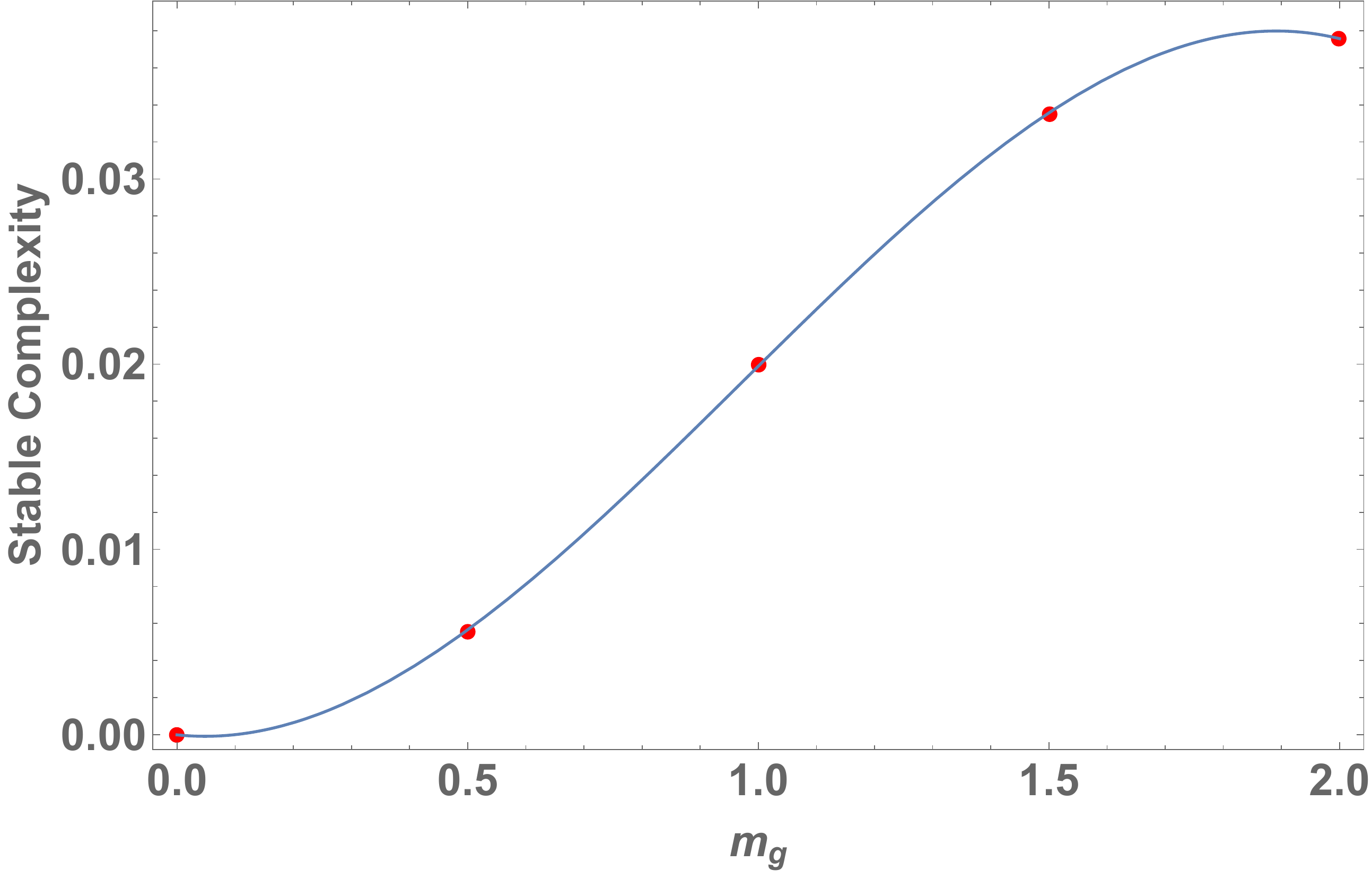}\ \\
  \caption{Left plot: The evolution functions of HC for different $m_g$.  The maximum value (middle plot) and the stable value (right plot) of the HC as the function of $m_g$. Here we have set $m=1$,$q=0$, $v_0=0.01$ and $l = 2$.}
 \label{fig:Cvst}
\end{figure}

After obtaining the HRT surface $\gamma_{\mathcal{A}}$, we can then work out the codimension-one surface $\Gamma_{\mathcal{A}}$, which characterizes the subregion complexity bounded by the HRT surface $\gamma_{\mathcal{A}}$ as well as boundary area $\mathcal{A}$, which is exhibited in Fig.\ref{fig:3dzvxC}. We present the evolution of the HC for different $m_g$ in Fig.\ref{fig:Cvst}.
One could see that, for different values of $m_g$, the evolution has a common feature that, at the early stage, the HC rises as the time evolves and then arrives at the maximum value. After that, it quickly drops and reaches to a stable value at the final stage.

To quantitatively explore what role $m_g$ plays, we plot the maximum value and the stable value of the HC as the function of $m_g$ in Fig.\ref{fig:Cvst}.  One can observe that with the increase of $m_g$, the maximum value of HC decreases, while the stable value at later stage increases, which is in contrary with the case of HEE. In addition, the HC with large values of $m_g$  takes shorter time to achieve stability, than with the case with smaller $m_g$, which is also consistent with the evolutions of the $z_*$ and the HEE displayed above.

\begin{figure}[ht!]
 \centering
  \includegraphics[width=5cm]{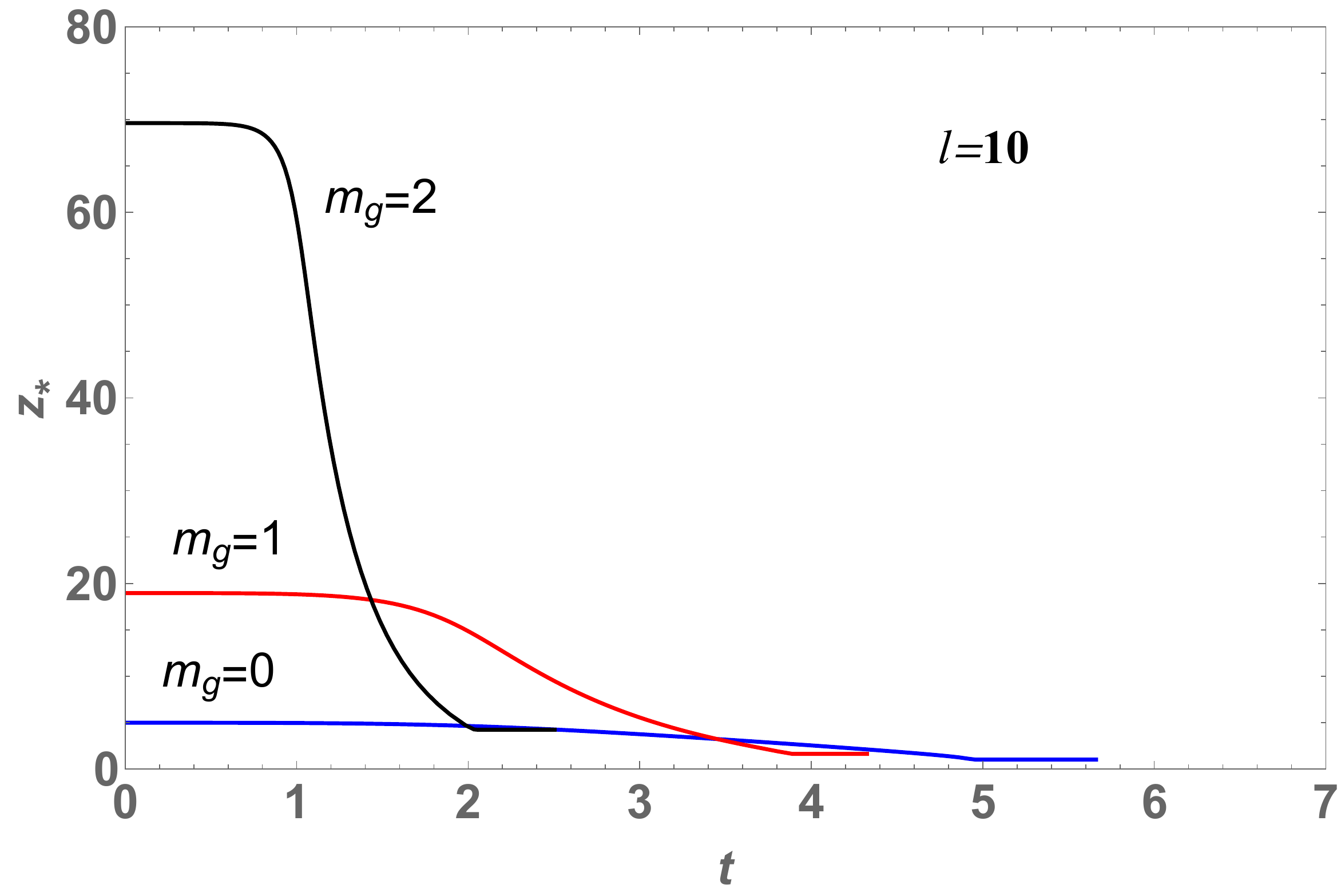}\ \hspace{0.1cm}
  \includegraphics[width=5cm]{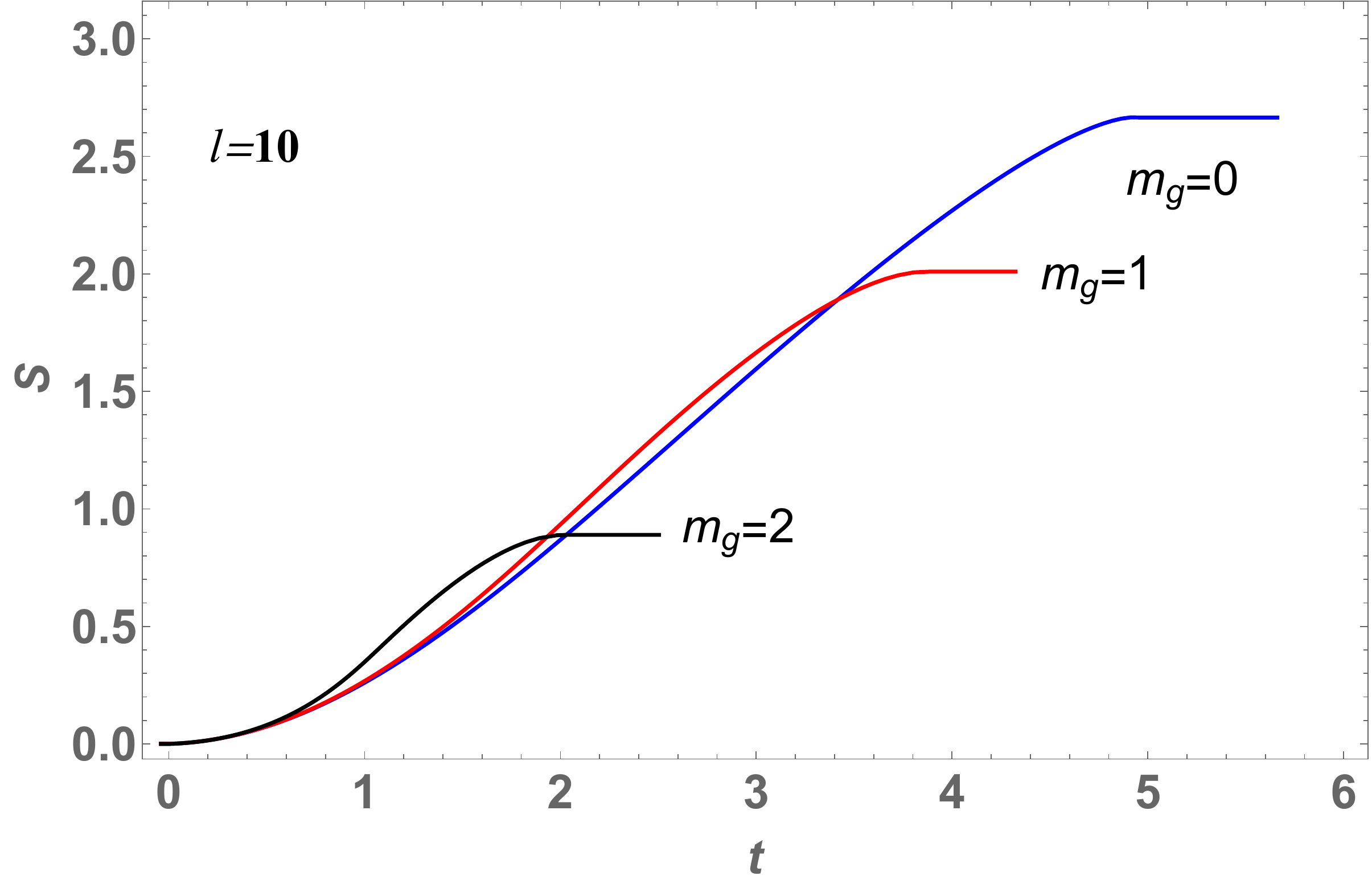}\ \hspace{0.1cm}
  \includegraphics[width=5cm] {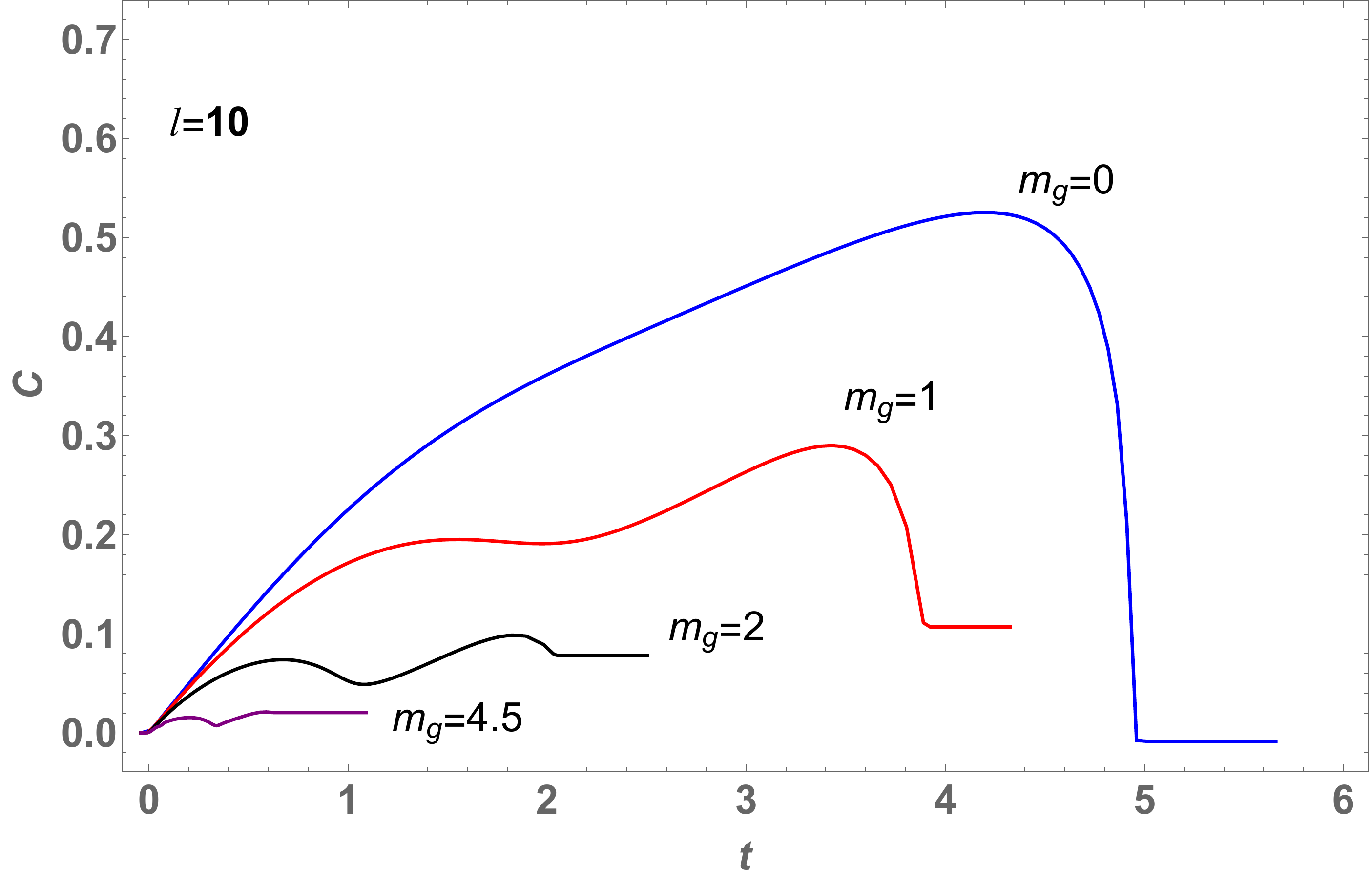}\
	  \caption{Left plot: The evolution of $z_{*}$ versus $m_g$.
	  Middle plot: The evolution of the HEE versus $m_{g}$. Right plot: The evolution of the HC versus $m_g$. Here we have set $l$=10, $q=0$, $m=1$ and $v_{0}=0.01$.}
 \label{fig:ml10ztst}
\end{figure}

We then study the effect of $m_g$ on HEE and HC with larger size of the subregion, i.e, $l=10$. The results are shown in Fig.\ref{fig:ml10ztst}. We  see that for larger width of the strip ($l=10$),
the effect of $m_g$ on the turning point $z_{*}$ and HEE are just very similar to the case with small widths.

However, novel properties are observed for HC with bigger size, $l$, (see the right plot of  Fig.\ref{fig:ml10ztst}). One could see that, as $m_g$ increases,  before the final drops of HC, two peaks are emerged. This behavior is truly different from the case with $l=2$ where only one peak has been observed. Note that as $m_g$ increases, the stable value of HC approaches the second peak, and also HC does not fall from the second peak but just directly reaches to the stable final region. This phenomena is a novel observation and the deep physical explanation is called for.

\begin{figure}[ht!]
 \centering
  \includegraphics[width=7.9cm]{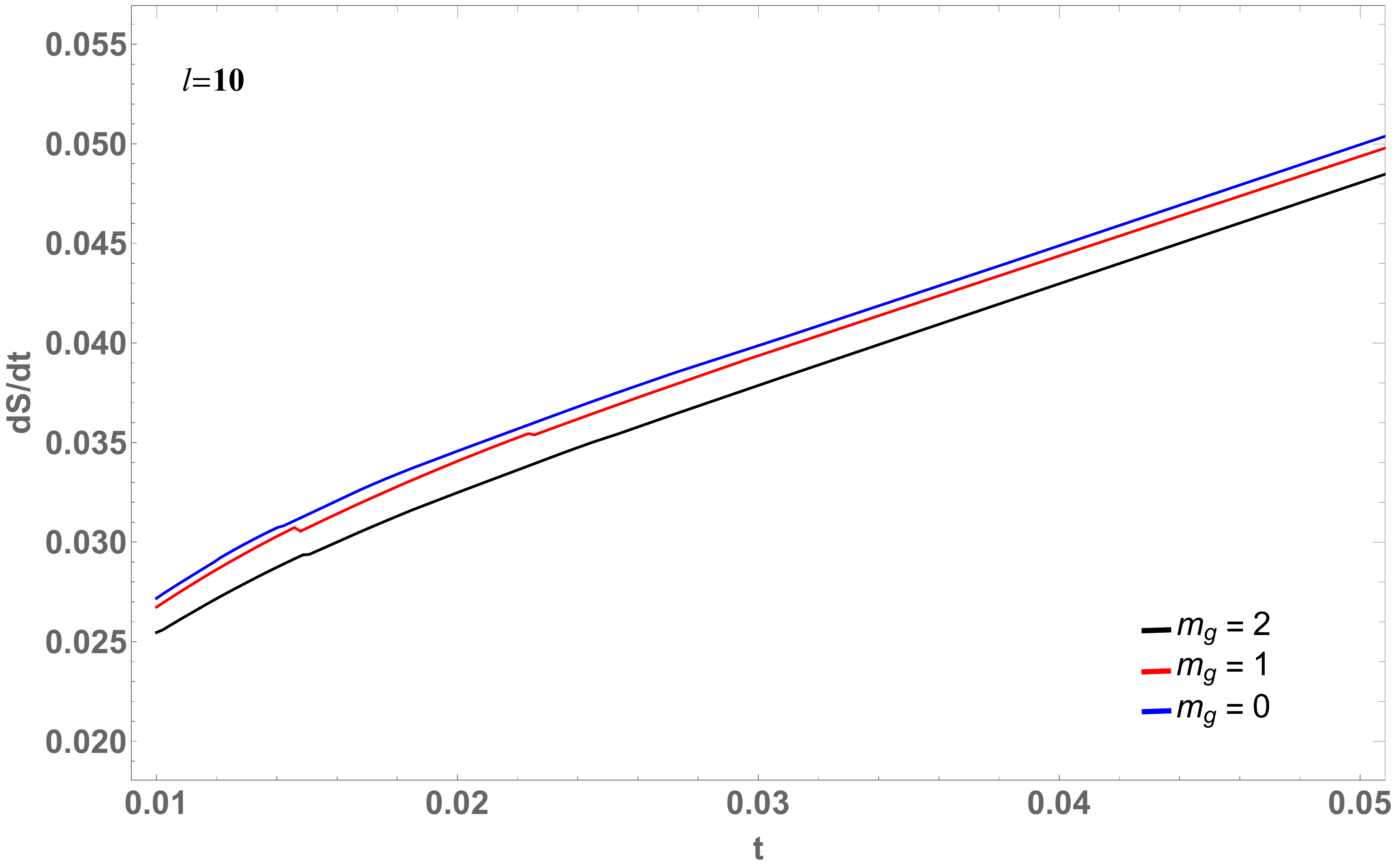}\hspace{0.1cm}
  \includegraphics[width=7.9cm]{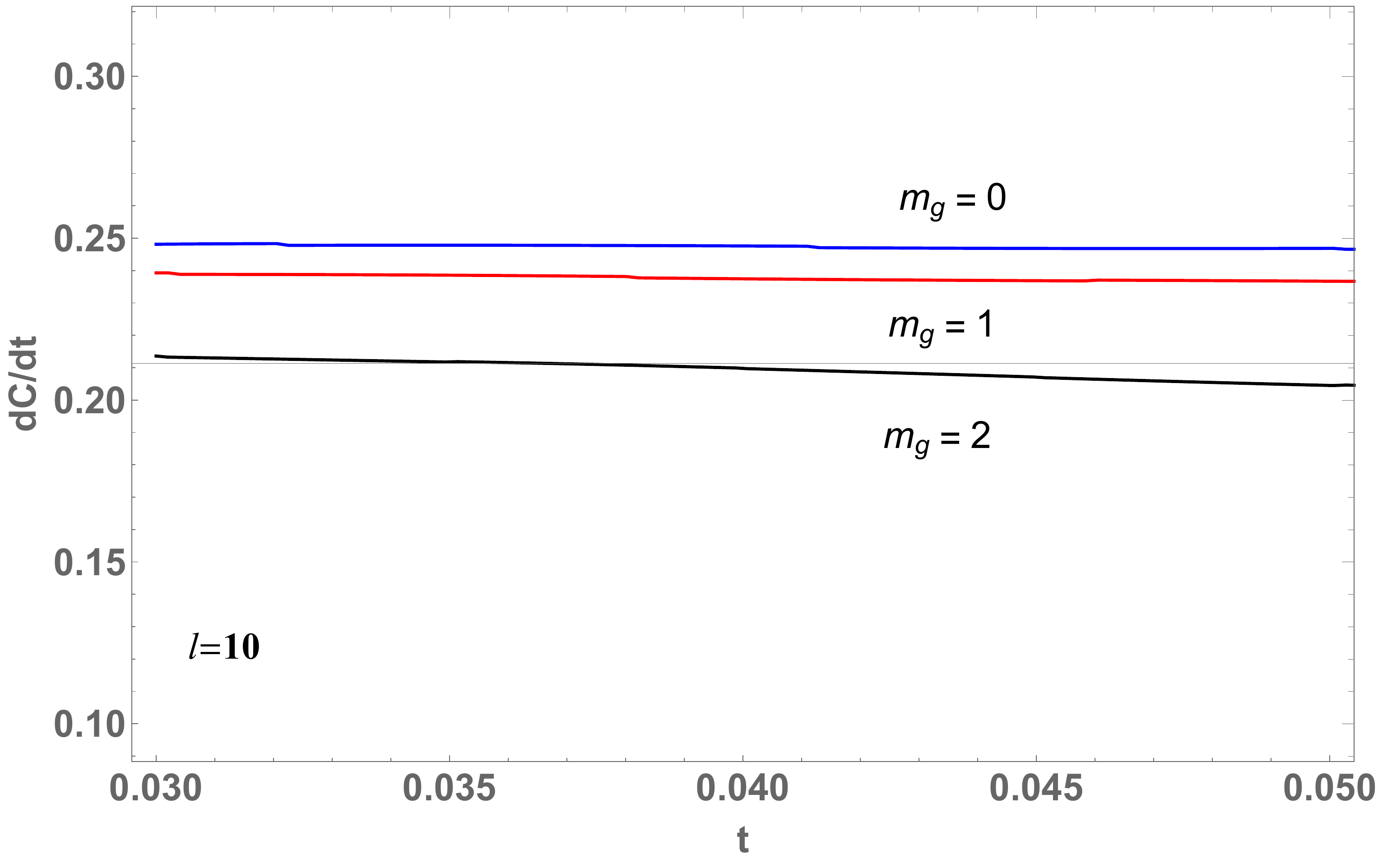}
	  \caption{Left plot: the growth of HEE for different $m_{g}$.
	  Right plot: the evolution of HC for different $m_{g}$.
	  Here we have set $q=0$, $m=1$, $v_{0}=0.01$ and $l=10$.}
 \label{fig:mCgrowth}
\end{figure}

Another interesting phenomena observed at the early stage is that, the evolution of both HEE and HC seems to slightly depend on the parameter $m_g$. In order to explicitly show the effect of $m_g$ on the initial evolution, we study the growth of HEE and HC for the size $l=10$, of which the result is shown in Fig.\ref{fig:mCgrowth}\footnote{Due to the numerical precision, the result for the growth of HC can not start from the initial time $t=0.01$.}.
From the figures, one could see that both growth functions are almost linear with respect to time. Also, the bigger graviton mass decreases the rate of growth of both HEE and HC. This result is reasonable since adding the parameter $m_g$ to the system is equivalent to introducing momentum relaxations in the boundary CFT theory.
We then  do the parallel computations for different values of $l$.  The result is that the linear behavior does not depend on the size of the system $l$.

\subsection{HEE and HC in charged BTZ black hole}\label{sub-charge}

\begin{figure}[ht!]
 \centering
  \includegraphics[width=5cm] {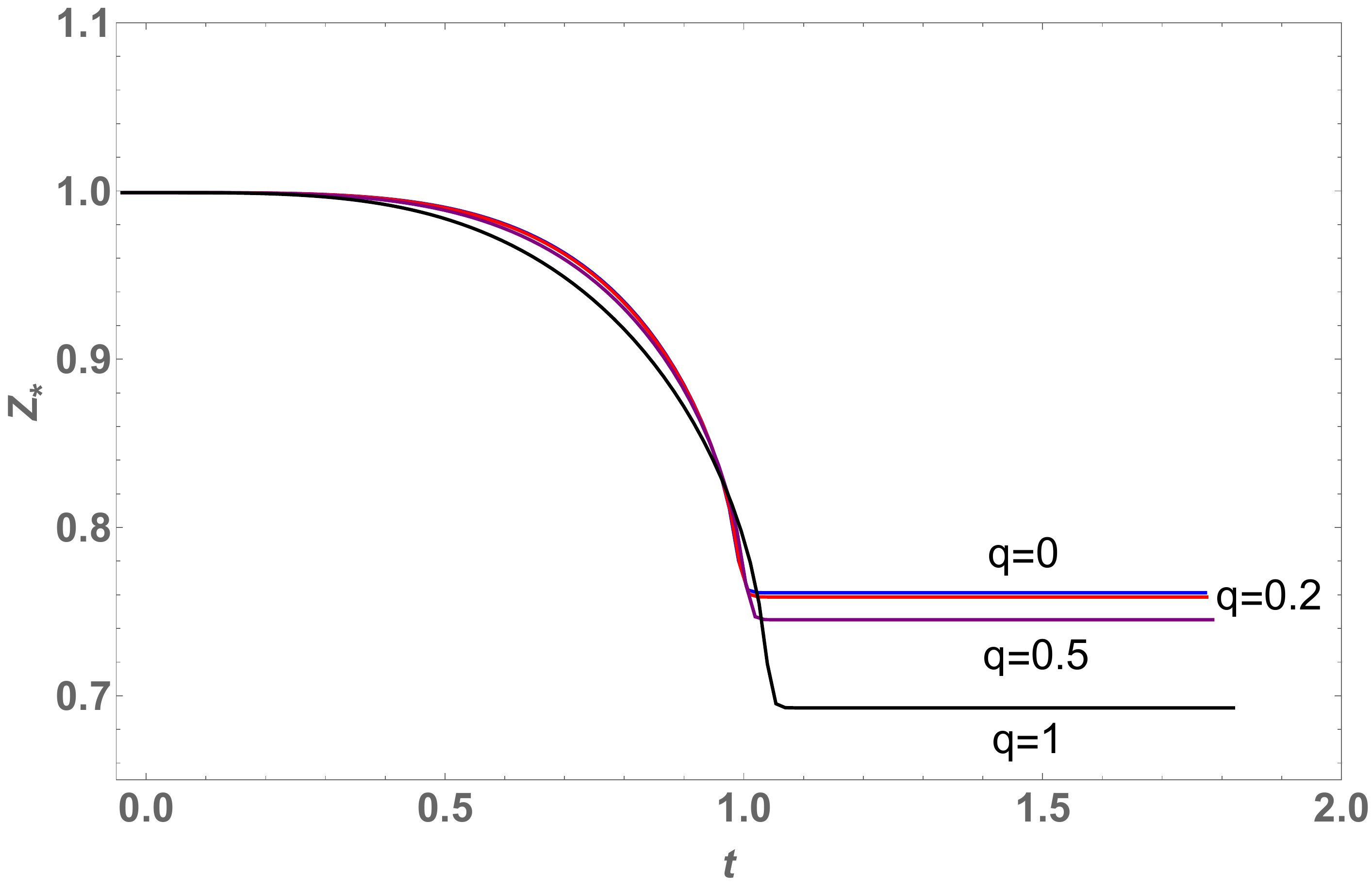}\ \hspace{0.1cm}
  \includegraphics[width=5cm] {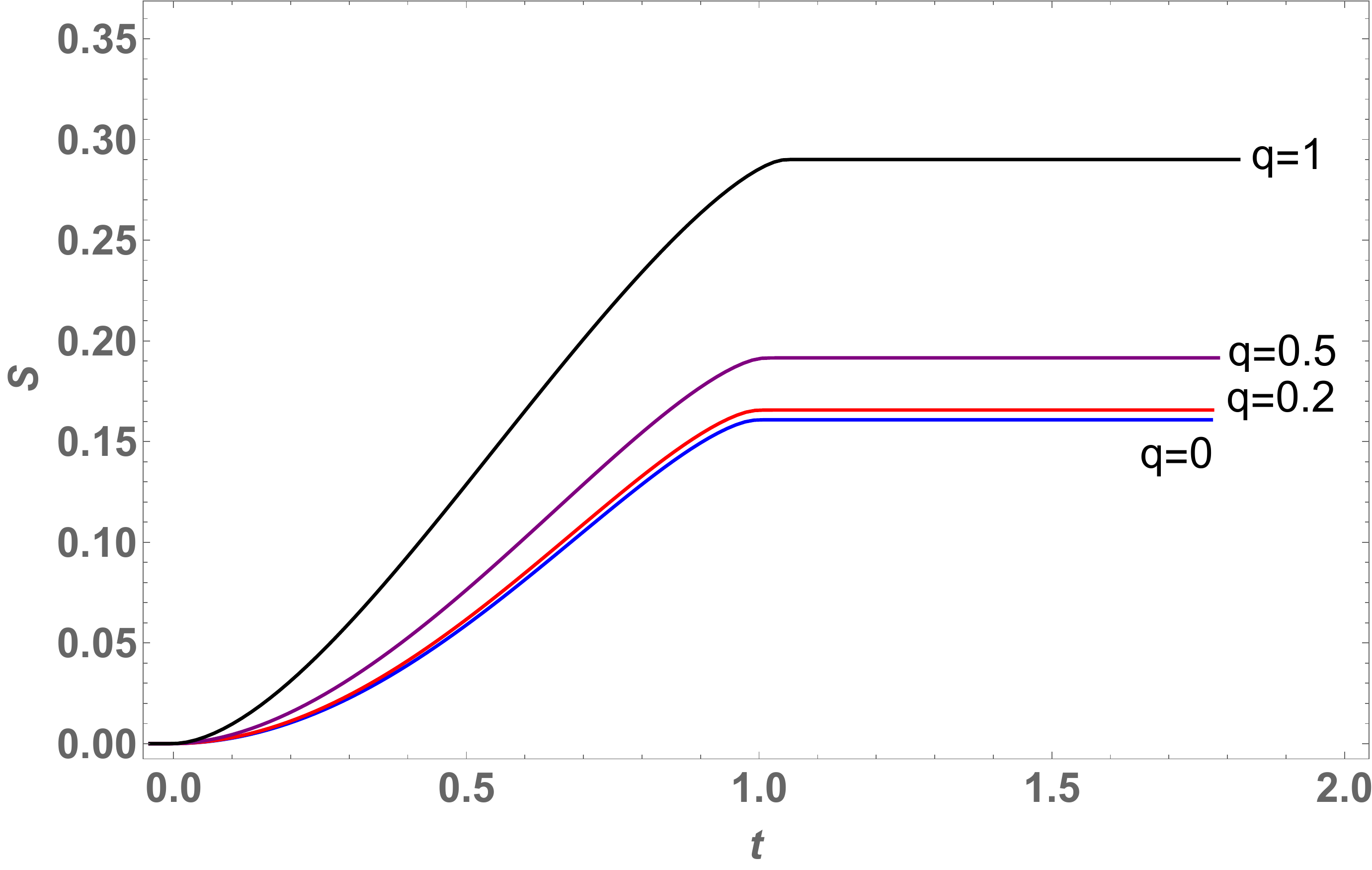}\ \hspace{0.1cm}
    \includegraphics[width=5cm] {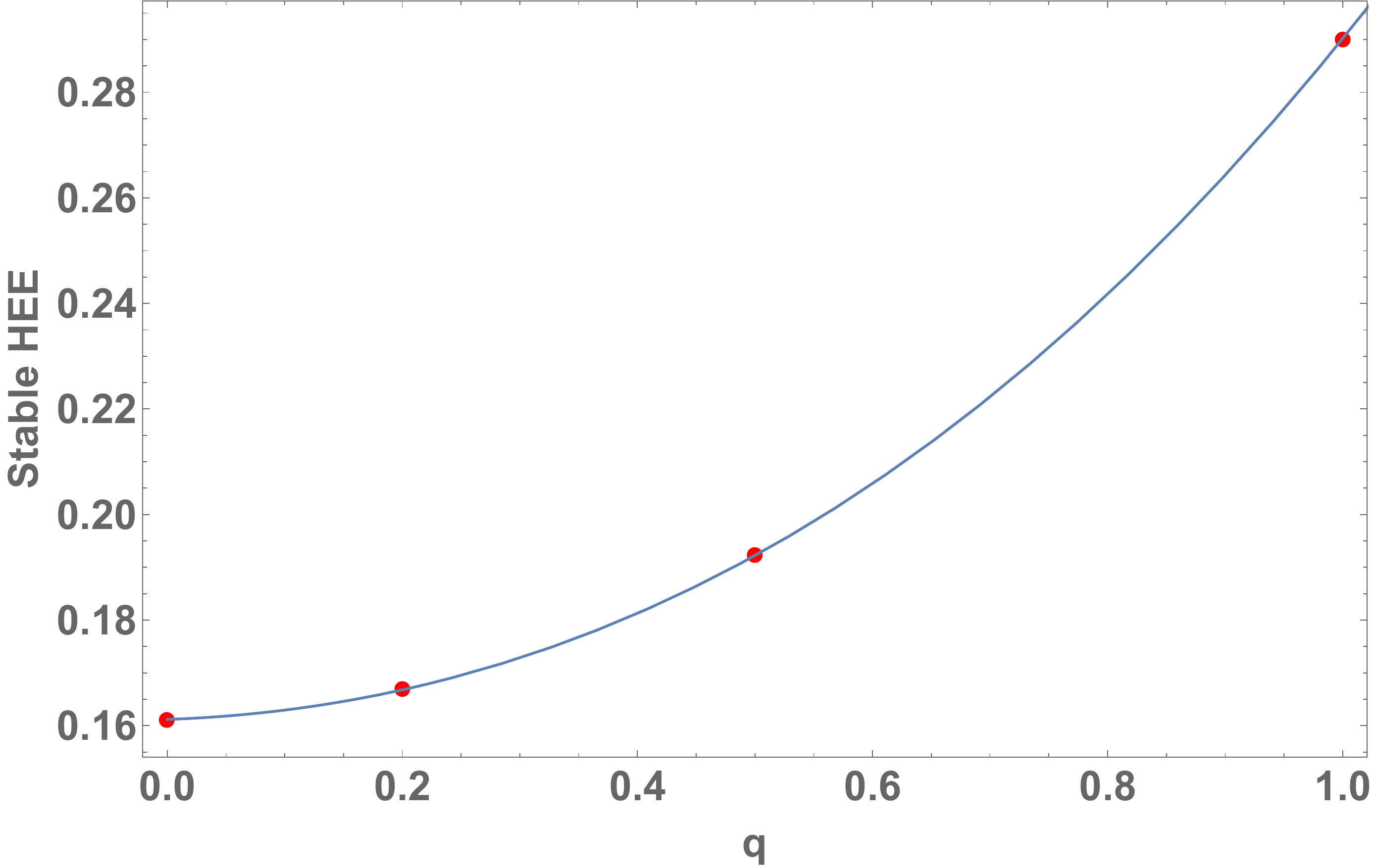}\\
  \caption{Left plot: The evolution of $z_{*}$ for different charge $q$.
  Middle plot: The evolution of the HEE for different charge $q$. Right plot: The relation between the stable value of the HEE at late time and the charge $q$.
  Here we have set $m=1$, $m_g=0$, $v_0=0.01$ and $l = 2$.}
 \label{fig:qSEEvsT}
\end{figure}

In this subsection, we study the effects of the charge $q$ on the evolutions of the HEE and HC. To this end, we turn off $m_g$, by setting $m_g=0$. As in the neutral case, we first fix the strip width as $l=2$. Also, we set $m=1$ and $v_{0}=0.01$.

Similarly, we study the evolution of the HRT surface $\gamma_{\mathcal{A}}$ in $(x,v,z)$ space and the projection in $(x,z)$ plane. We show the evolutions of the turning point $z_*$ and the HEE in Fig.\ref{fig:qSEEvsT}.
Comparing with the neutral massive gravity case, we summarize the properties of the charged case as follows:
\begin{itemize}
	\item At the early stages of the evolution, $z_*$ behaves almost same for different $q$ (see the left plot in Fig.\ref{fig:qSEEvsT}). However, this behavior is different from the the neutral massive gravity case shown in the left plot in Fig.\ref{fig:SEEvsT}. One could see that, the HEE vanishes at the early stage for various values of $q$, which  is similar to the case of neutral massive gravity. As we mentioned above, it is because we have subtracted the part of HEE which is dual to the AdS geometry.
	\item Then, one could notice that, both $z_{*}$ and HEE arrive at a stable stage finally, and the charge $q$ has a definite print on the final stable values. As $q$ increases, the turning point $z_*$ decreases but the HEE increases.
	Quantitatively, we show the relation between the stable value of the HEE and $q$ in the right plot of Fig.\ref{fig:qSEEvsT}.
	\item Another point we find is that, solutions with bigger charge $q$ is more difficult to saturate into equilibrium, which is denoted by the point that, for bigger charges, HEE needs longer times to become stable. This phenomena is the same behavior as in the case of other charged black holes. For those solutions also, one could see that charge always slows down the thermalization process, for instance the case in four dimensional background which has been addressed in \cite{Camilo:2014npa}.
\end{itemize}

\begin{figure}[ht!]
 \centering
  \includegraphics[width=5cm] {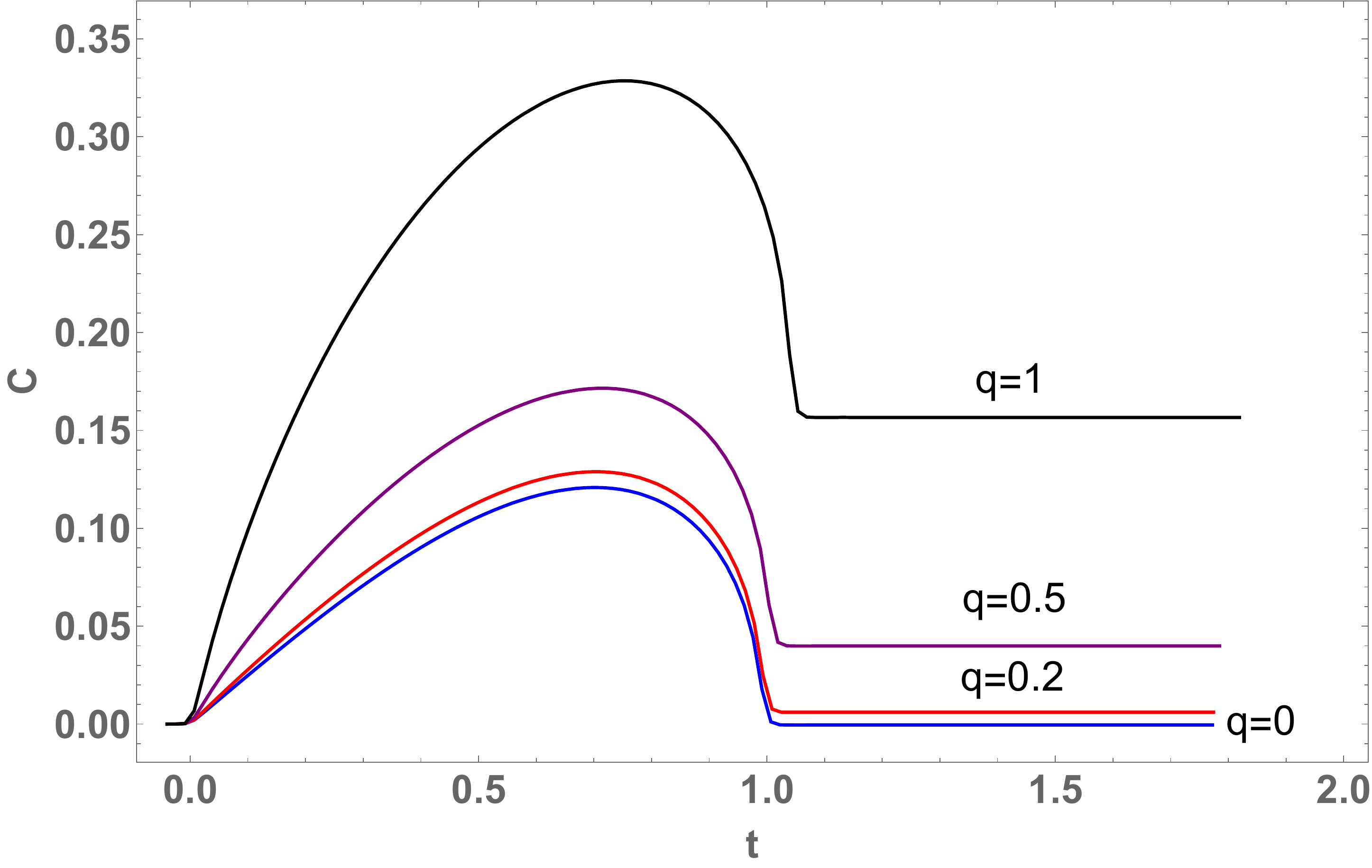}\ \hspace{0.1cm}
  \includegraphics[width=5cm] {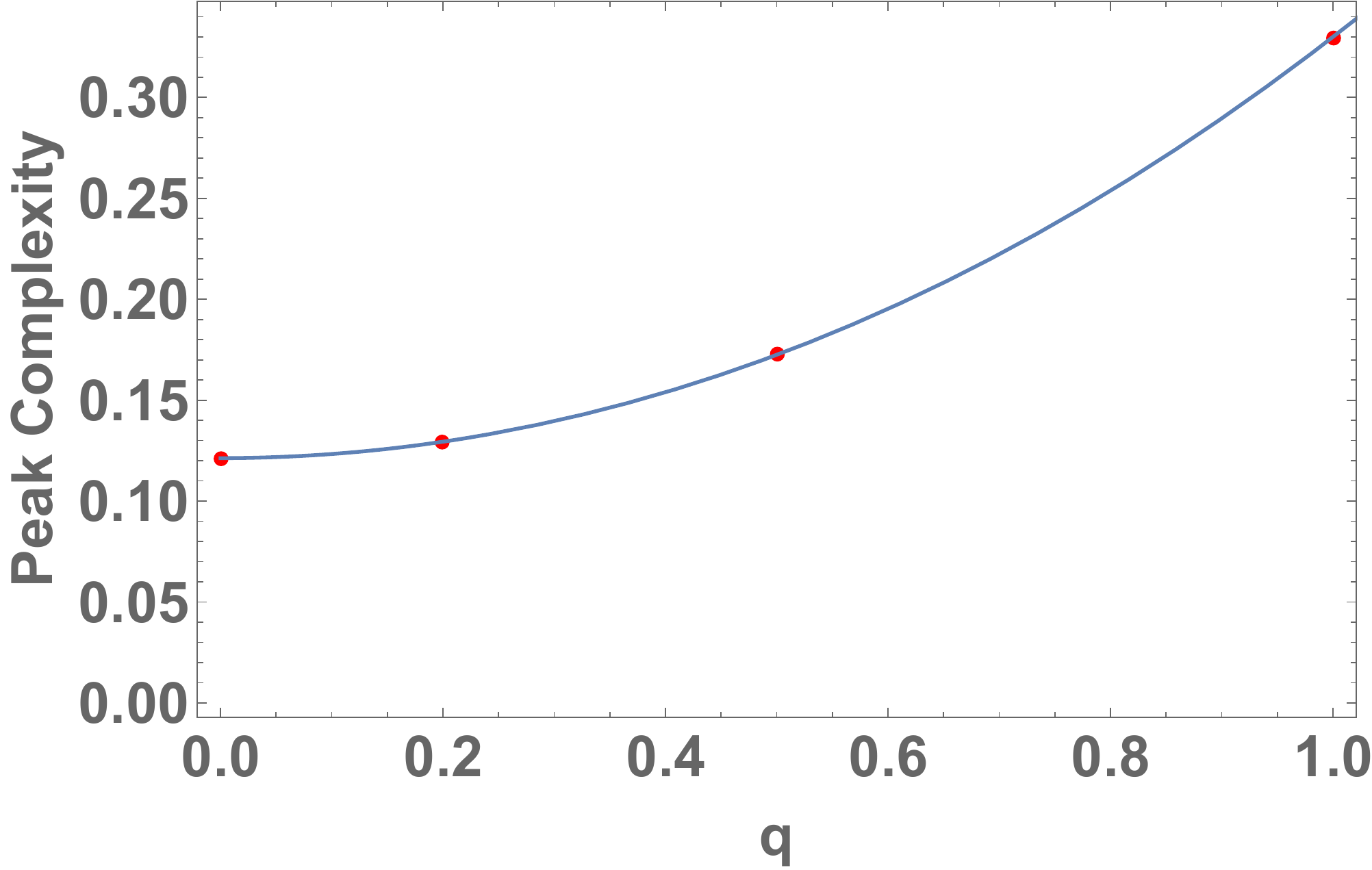} \ \hspace{0.1cm}
   \includegraphics[width=5cm] {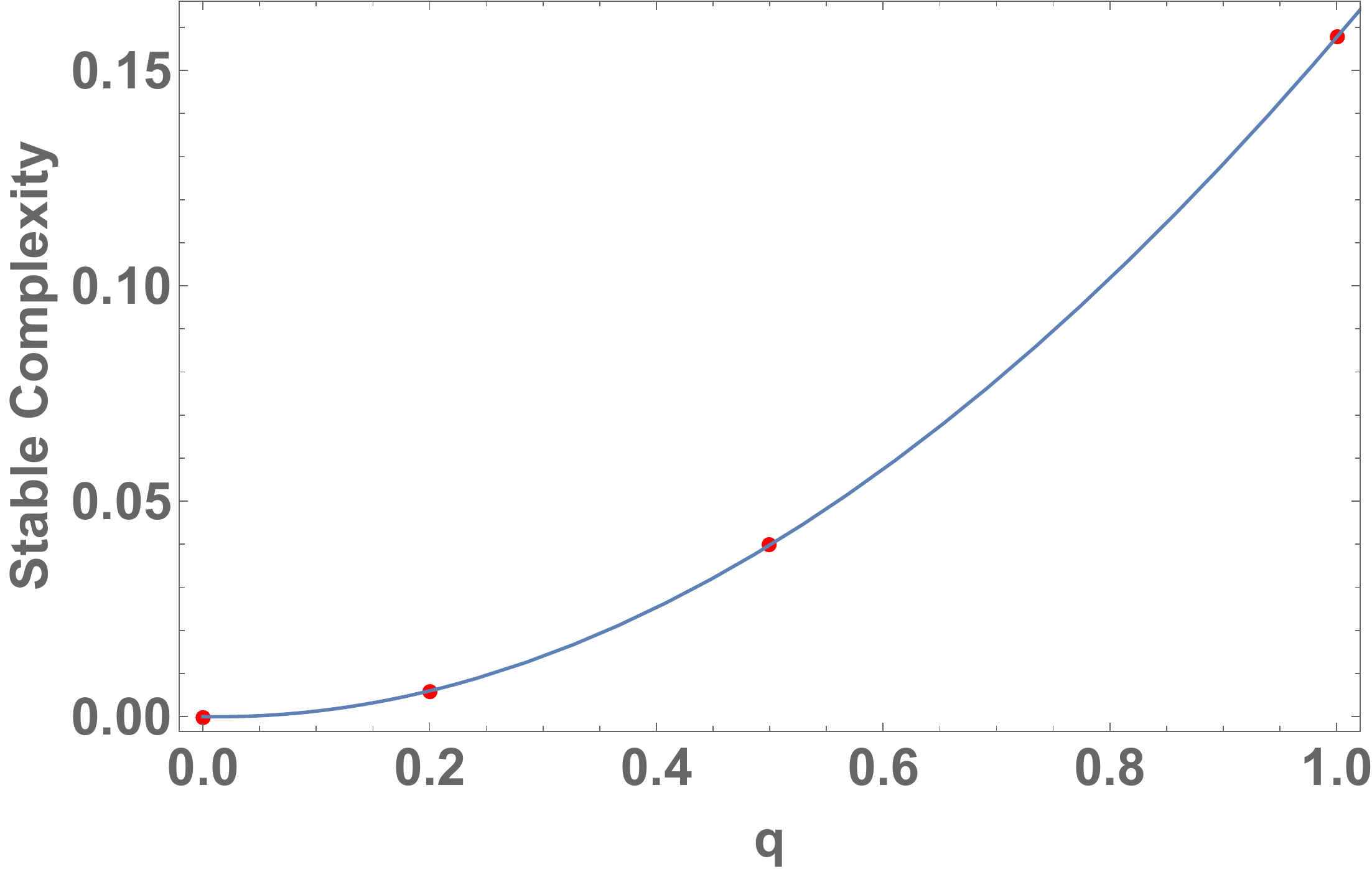}
  \caption{The evolution of the HC, for different $q$, is shown in the left plot, while the maximum value versus $q$ in the middle plot and the stable value versus $q$ in the right plot.
  Here, we have set $l=2$, $m=1$, and $v_{0}=0.01$.}
 \label{fig:qCvst}
\end{figure}

Now, we turn to study the HC in the background of charged BTZ black holes.
First, we present the evolution of the codimension-one surface $\Gamma_{\mathcal{A}}$, and then we plot the evolutions of the HC for different values of $q$ in Fig.\ref{fig:qCvst}.
This evolution behavior is similar to the case of neutral BTZ black hole (see Fig.\ref{fig:Cvst} or reference \cite{Chen:2018mcc}). Additionally, for bigger charges, similar to the behavior of $z_*$ and HEE, HC  also takes longer times to arrive at the stable stage.

We also present the behavior of the maximum value and the stable value of the HC as the function of $q$ in Fig.\ref{fig:qCvst}. One could see that the two values increase as $q$ increases which intuitionally makes sense as charge would introduce more degrees of freedom for each gates and therefore could significantly increase complexity. This has also been noticed in the case of complexity of purification studied in \cite{Ghodrati:2019hnn}.

\begin{figure}[ht!]
 \centering
  \includegraphics[width=7.4cm] {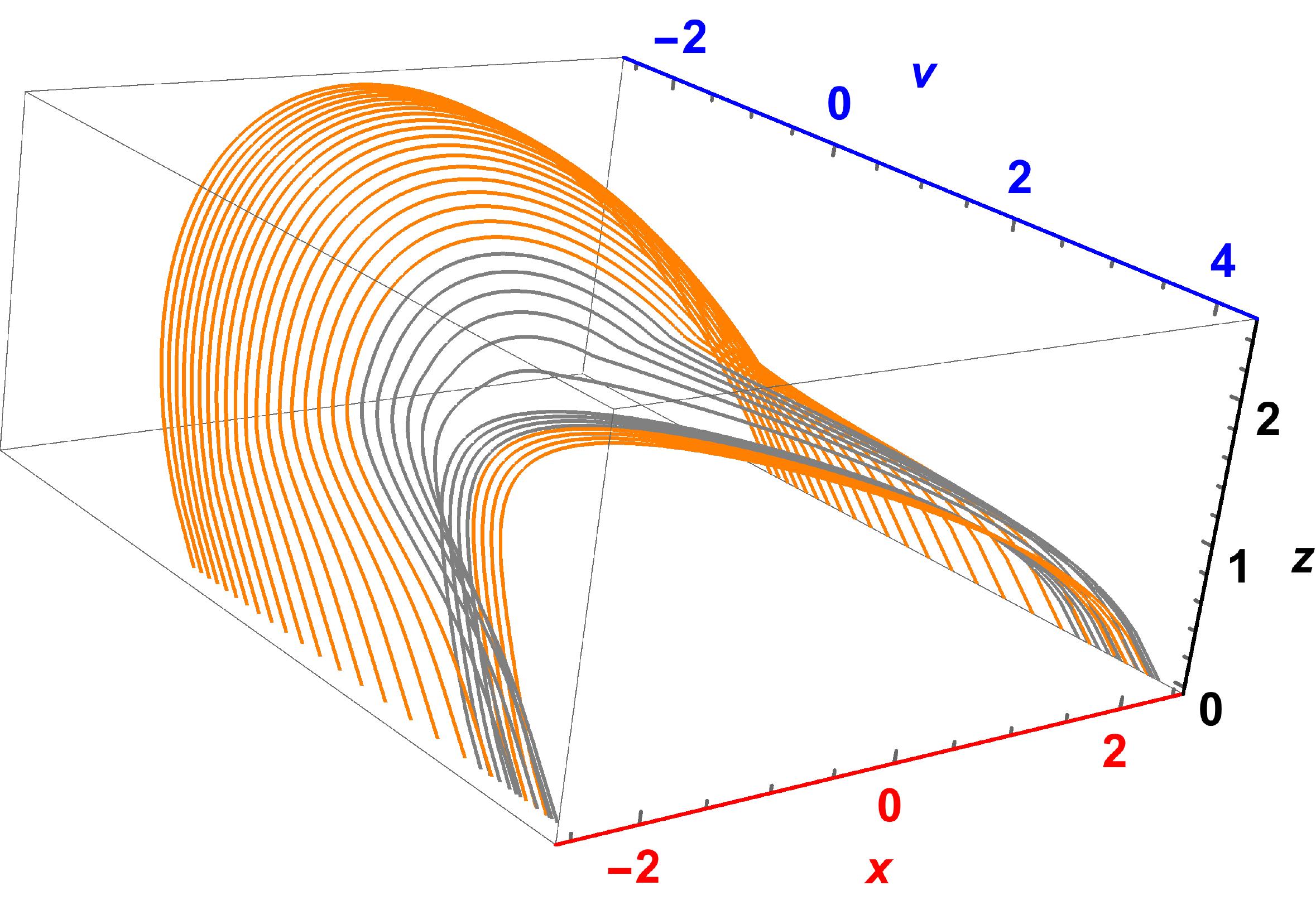} \ \ \ \ \ \ \
   \includegraphics[width=7.4cm] {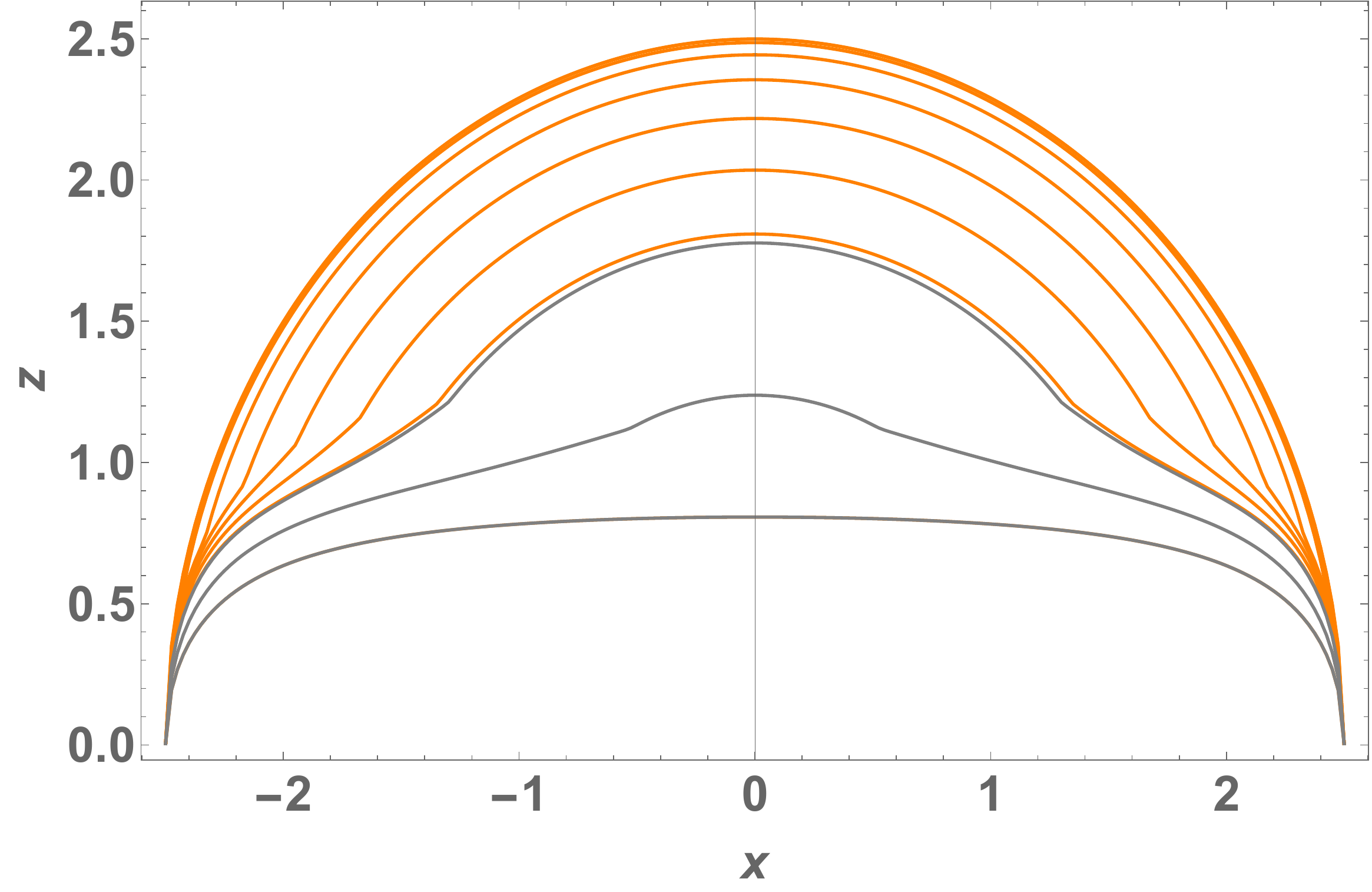}
  \caption{The evolution of HRT surface $\gamma_{\mathcal{A}} = (\tilde{z}(x), \tilde{v}(x))$ in the background of charged BTZ black hole.
  We have fixed $m=1$, $q=1.5$, $v_{0}=0.01$ and $l=5$ here.
  The left pattern shows the evolution in the $(x, v, z)$ space.
  The right pattern shows their projection in the $(x, z)$ plane.
  The HRT surface evolves from left to right in the left pattern and from top to bottom in the right pattern.}
 \label{fig:ql53dzvxSEEandql5zxSEE}
\end{figure}
\begin{figure}[ht!]
 \centering
  \includegraphics[width=7.4cm] {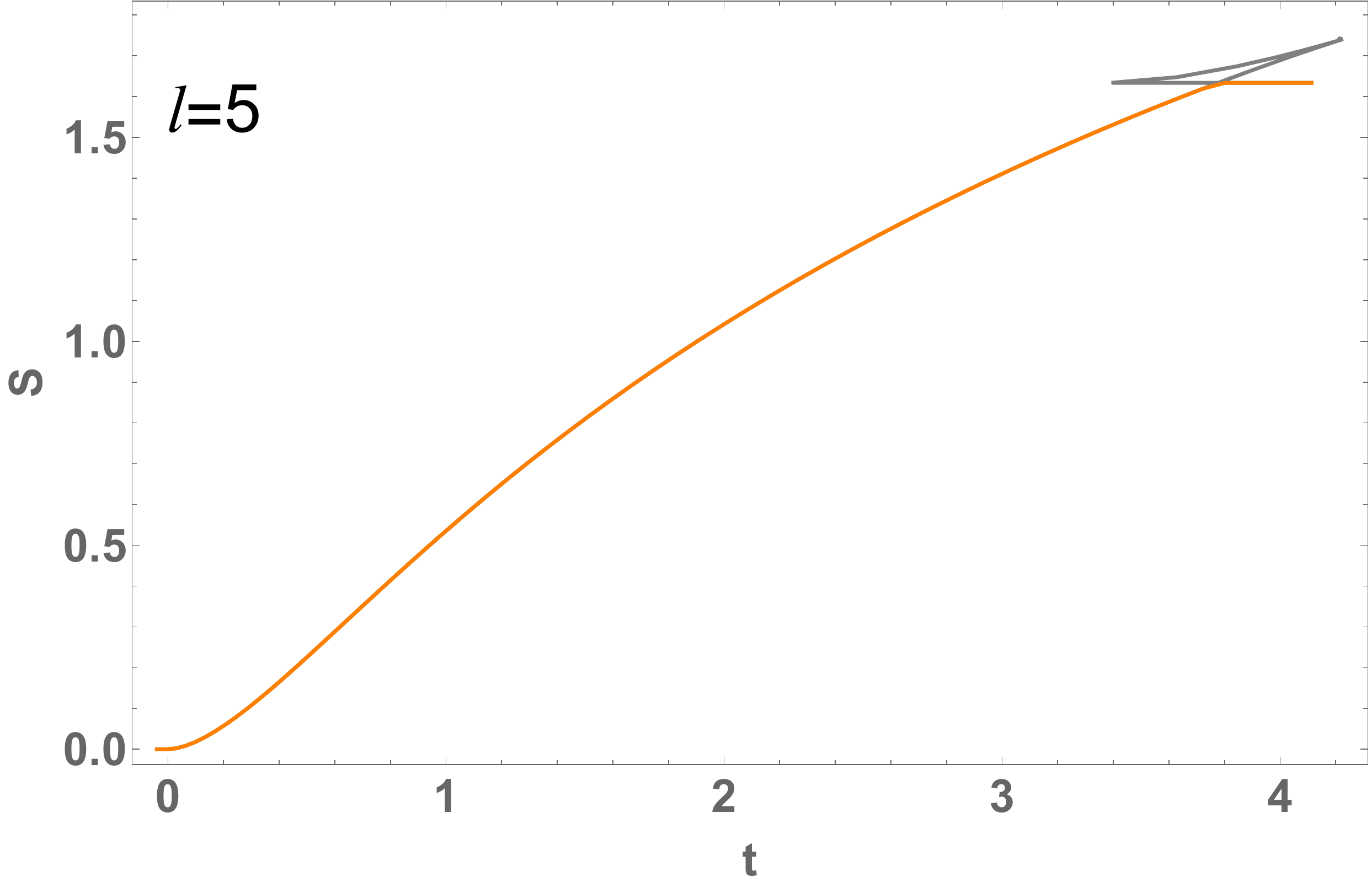}\ \hspace{0.1cm}
   \includegraphics[width=7.4cm] {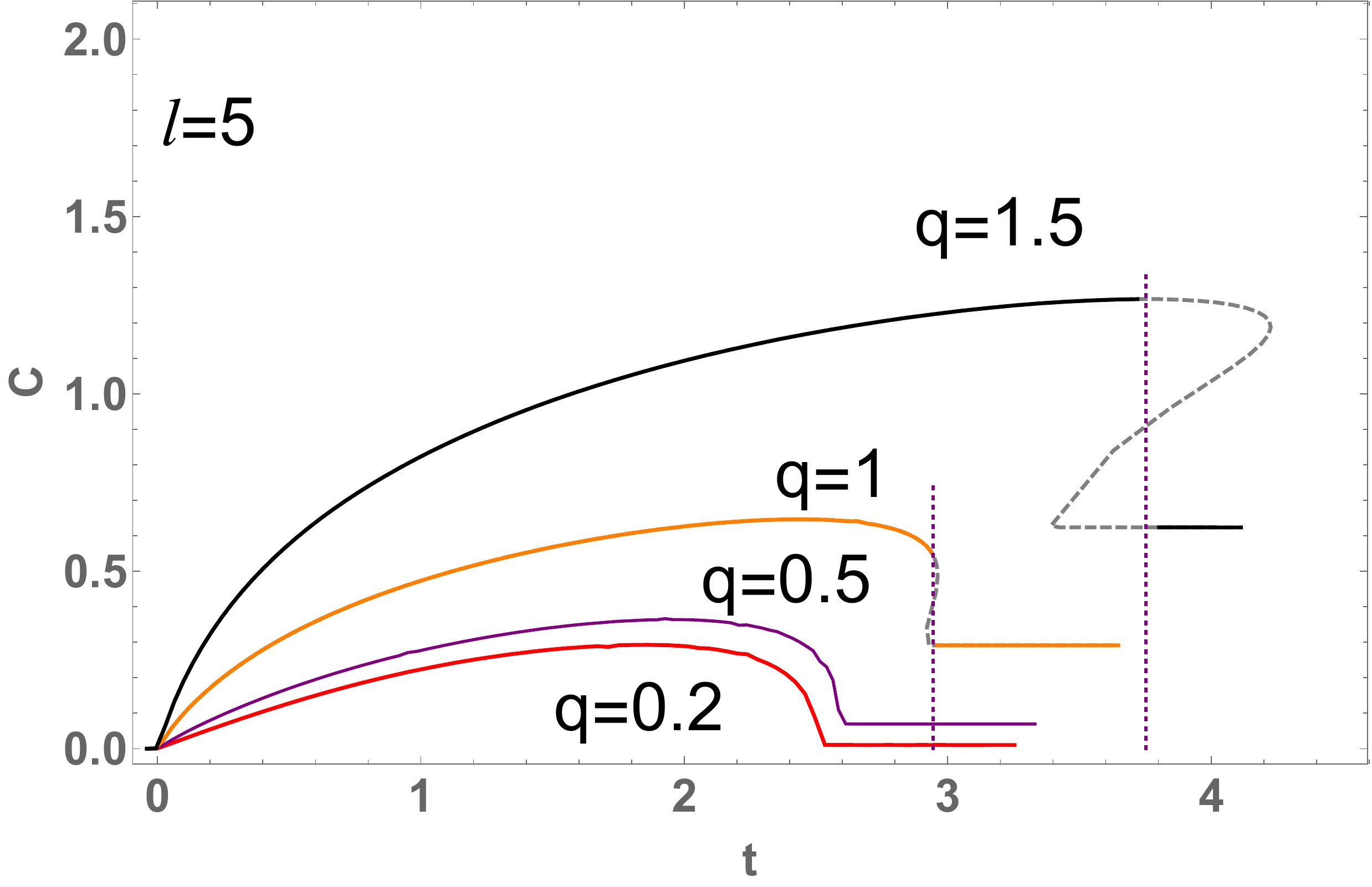}\\
     \includegraphics[width=7.4cm]{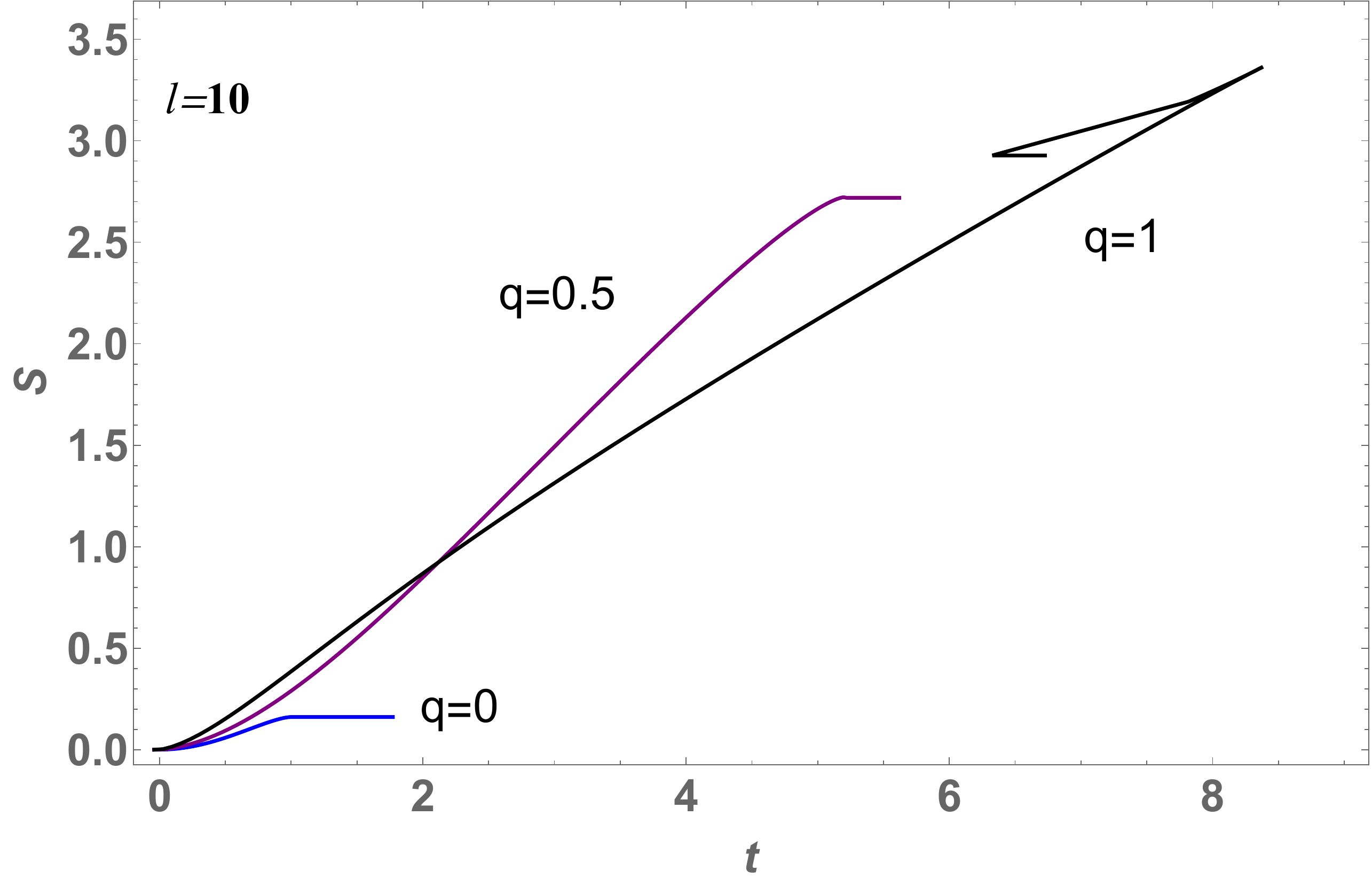}\ \hspace{0.1cm}
       \includegraphics[width=7.4cm] {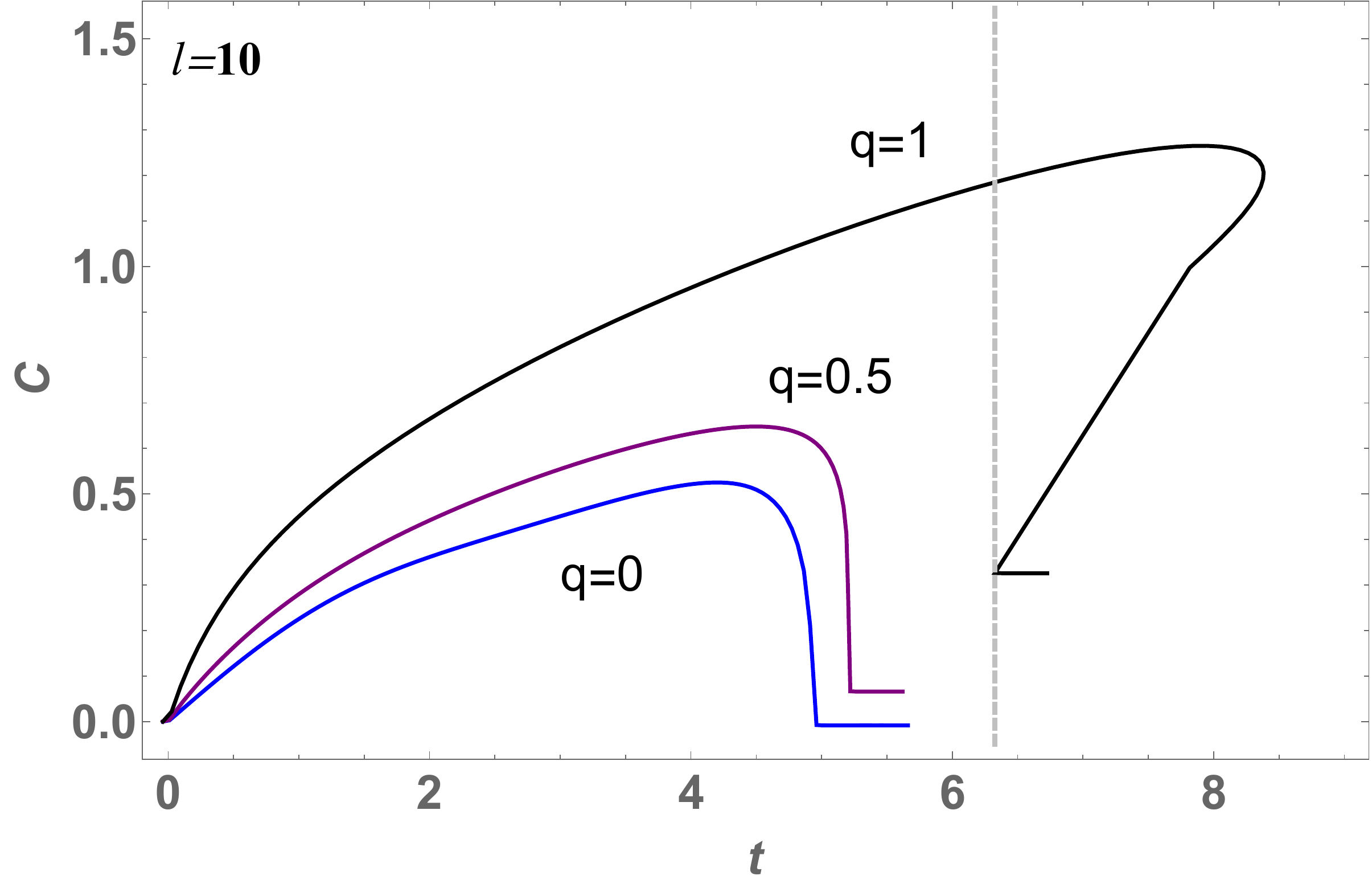}
  \caption{Left plots: The evolution of the HEE.
  Right plots: The evolution of the HC. In the upper panel, we set $l = 5$, while in the bottom panel we set $l = 10$. Here we have also set $m=1$, $m_g=0$, $v_0=0.01$.
  }
 \label{fig:qcl5andql5SEEvsT}
\end{figure}

\begin{figure}[ht!]
 \centering
  \includegraphics[width=7.4cm] {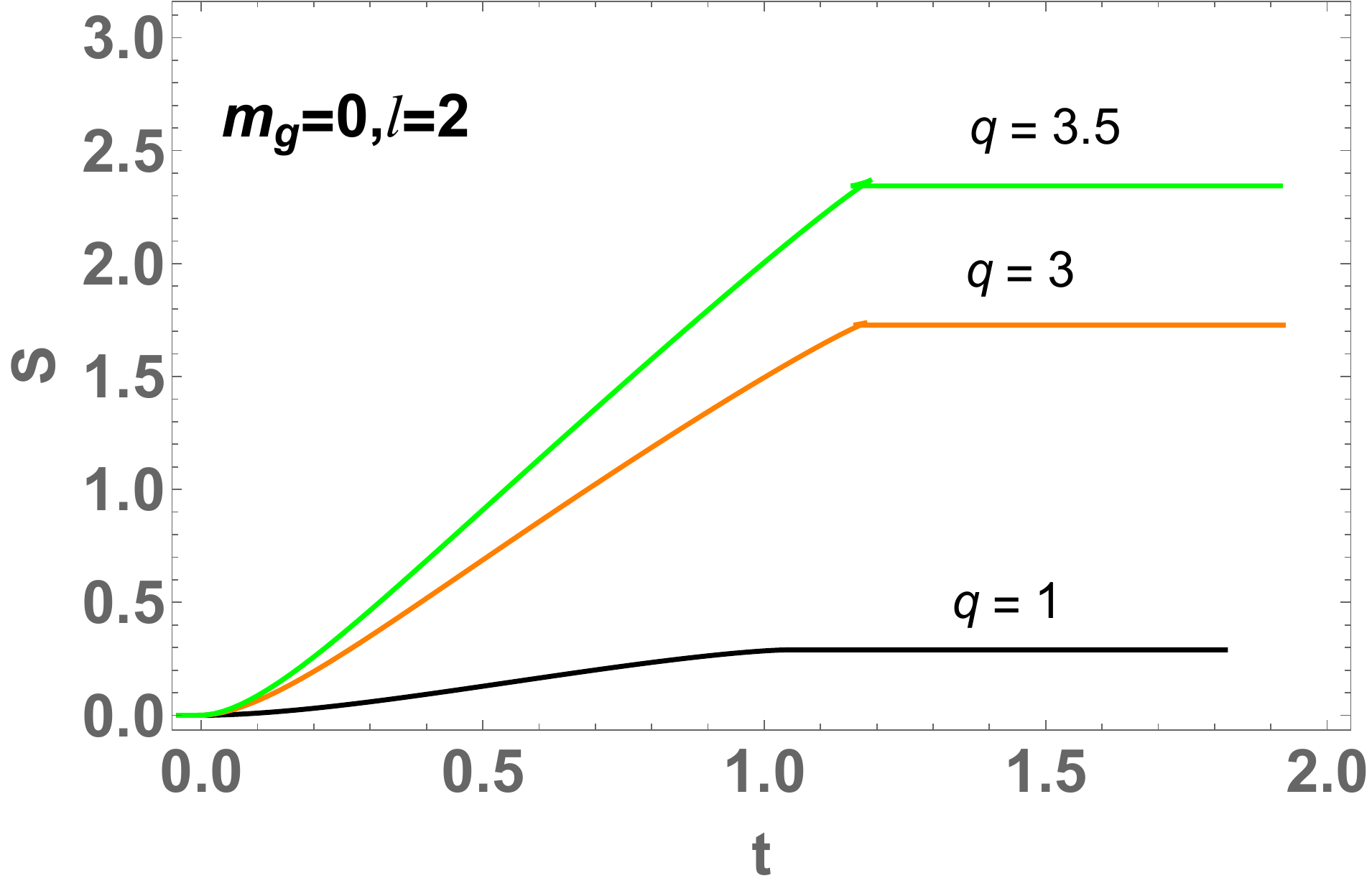} \ \ \ \ \ \ \
   \includegraphics[width=7.4cm] {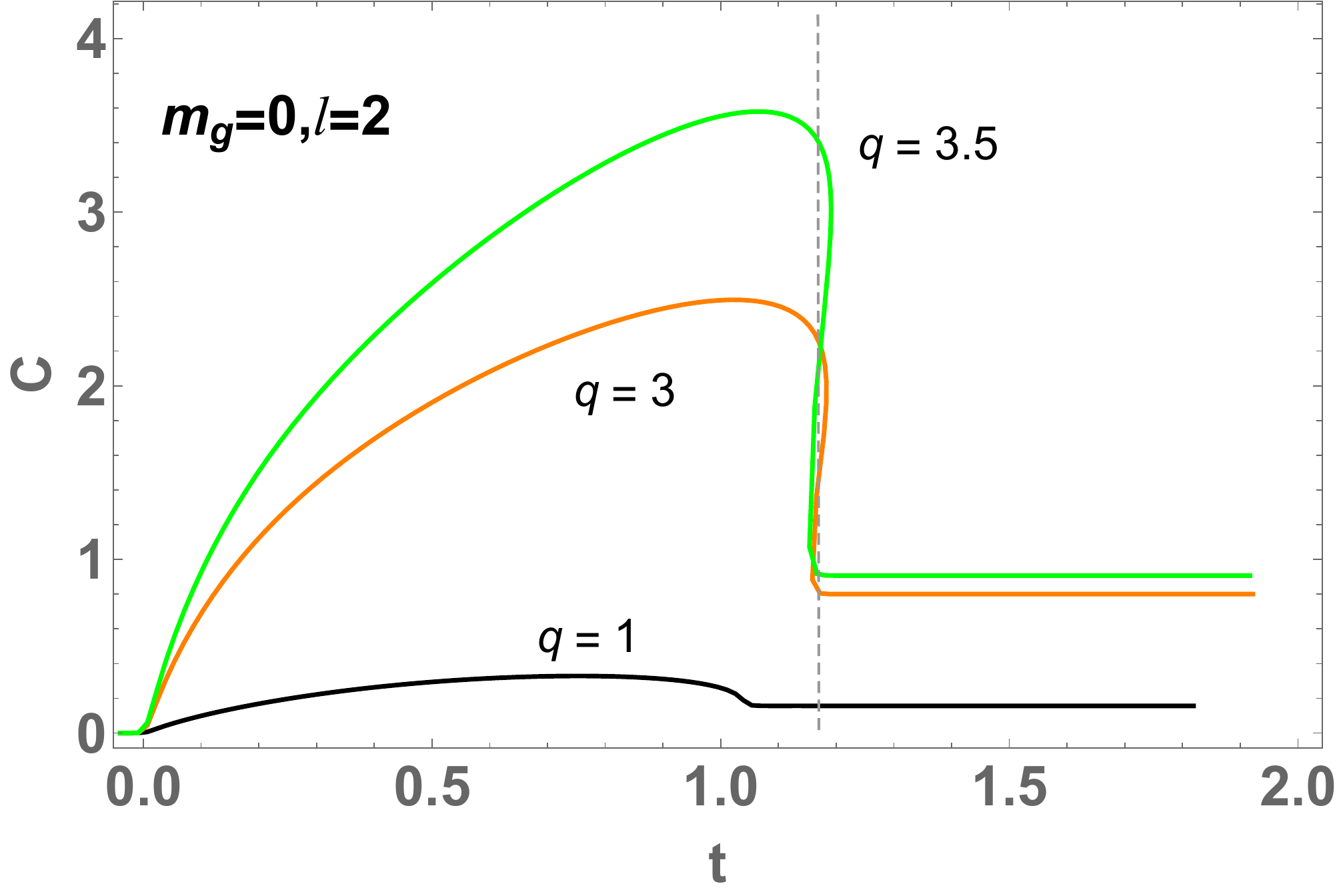}
  \caption{The evolution of the HEE (left plot) and the HC (right plot) in charged BTZ black hole for different $q$.
   Here we have set $m=1$, $m_g=0$, $v_0=0.01$ and $l = 2$.}
 \label{fig:mg0l2STCT}
\end{figure}

In Fig.\ref{fig:ql53dzvxSEEandql5zxSEE} we show the evolution of HRT surface $\gamma_{\mathcal{A}} = (\tilde{z}(x), \tilde{v}(x))$ in the background of charged BTZ black hole for the width of strip set as $l=5$. From this figure, it is obvious that the evolution of $\gamma_{\mathcal{A}}$ is no longer a continuous function, which is different from the case of the small width strips.
The discontinuous evolution is due to the jump in the minimal area surface, which corresponds to the swallow tail (gray line in left up plot in Fig.\ref{fig:qcl5andql5SEEvsT}) of the HEE
and therefore the multi-valued region (gray line in right up plot in Fig.\ref{fig:qcl5andql5SEEvsT}) for the HC. In fact, even for the size $l=2$, with big enough charge $q$, the swallow tail of the HEE and the multivaluedness of the HC can emerge again (Fig.\ref{fig:mg0l2STCT}).
We also show in Fig.\ref{fig:qcl5andql5SEEvsT} the HEE and HC for $l=10$.

Note that generally, the swallow tail in the behavior of HEE or the multivaluedness of the HC  implies  the existence of multiple solutions for the partial differential equation (PDE) at a given time.
We could note that when $q$ is turned down, the multivaluedness of the HC disappears even for the larger width of the strip, (see right plot in Fig.\ref{fig:qcl5andql5SEEvsT}). This behavior has also been observed in \cite{Ling:2018xpc}.

It is worthwhile to emphasize that the swallow tail in the HEE and the multi-values in the HC can only be emerged in more than three dimensional theories but not in the neutral AdS$_3$ theory. Even by increasing the parameters, such as the mass of black hole, to a very large point this behavior could not be emerged, as observed in the solutions of \cite{Chen:2018mcc}.
However, in our study, we find that, this feature can emerge in the charged AdS$_3$ theory with large enough $q$ and $l$. Moreover, comparing Fig.\ref{fig:qcl5andql5SEEvsT} and Fig.\ref{fig:mg0l2STCT}, we find that the bigger charge of black hole would promote the emergence of discontinuity with even smaller widths.

\begin{figure}[ht!]
 \centering
  \includegraphics[width=7.9cm]{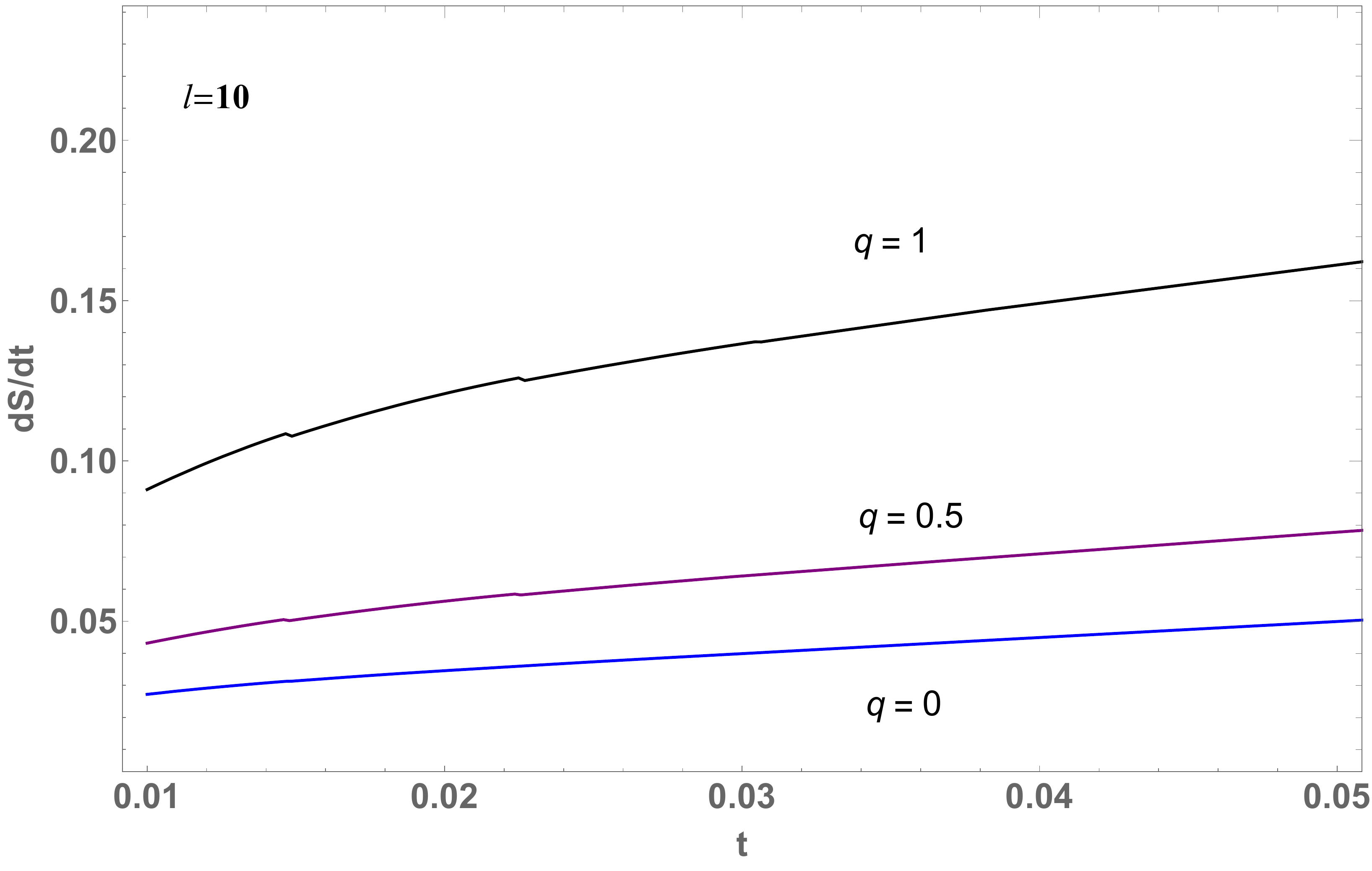}\hspace{0.1cm}
  \includegraphics[width=7.9cm]{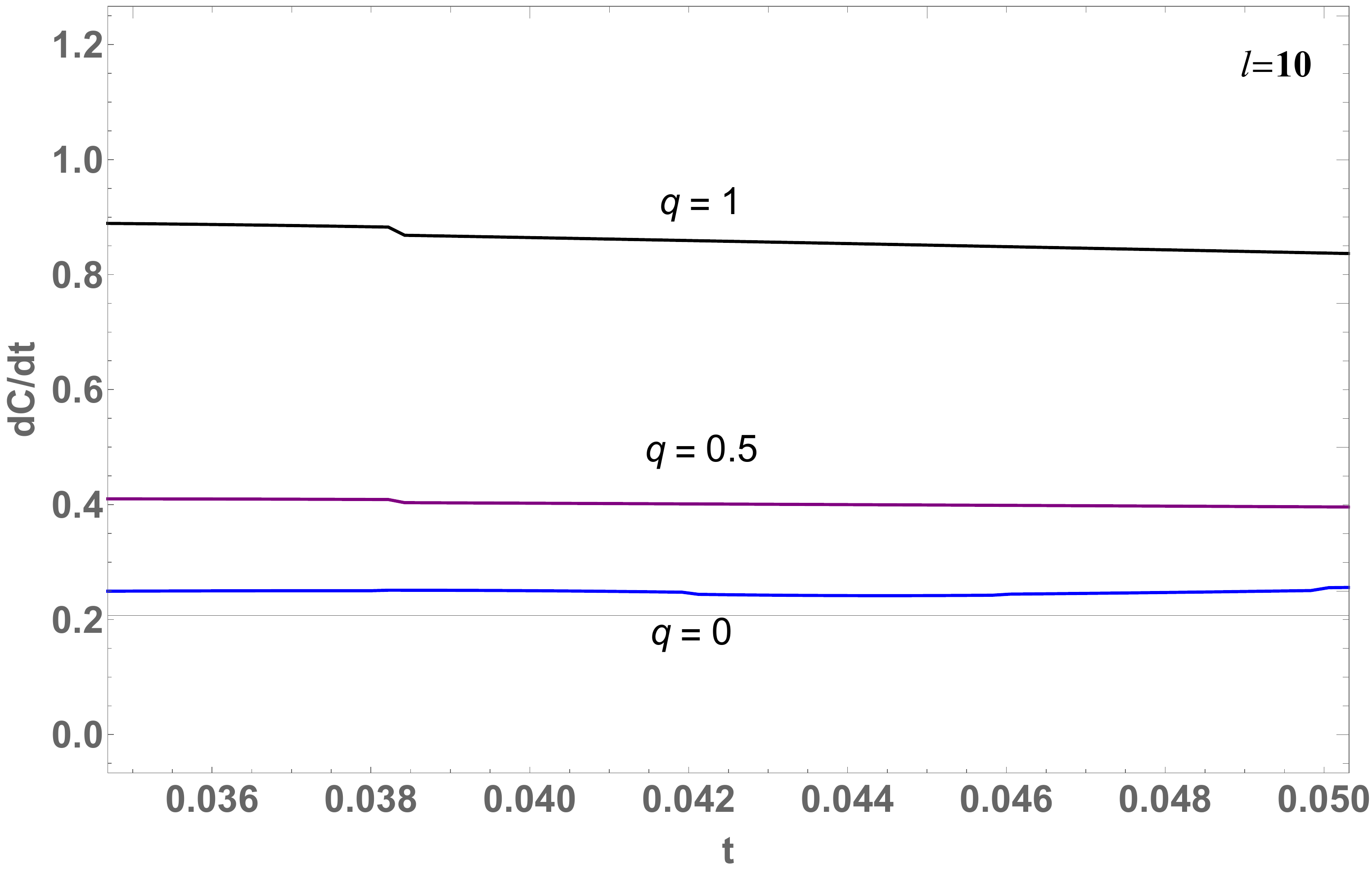}
	  \caption{Left plot: growth rate of HEE for different $q$.
	  Right plot: growth rate of HEE versus $q$.
	  Here we have set $m_g=0$, $m=1$, $v_{0}=0.01$ and $l=10$.}
 \label{fig:qSgrowth}
\end{figure}

Similarly, we study the effects of charge on the evolution of HEE and HC growth for large $l$ which is shown in Fig.\ref{fig:qSgrowth}. One could see the behavior is almost linear. Also, bigger charges  produces higher growth rates in both HEE and HC, which is similar to the effect of the graviton mass.
\subsection{HEE and HC in massive charged BTZ black hole}\label{sub-massive-charge}

\begin{figure}[ht!]
 \centering
  \includegraphics[width=7.8cm]{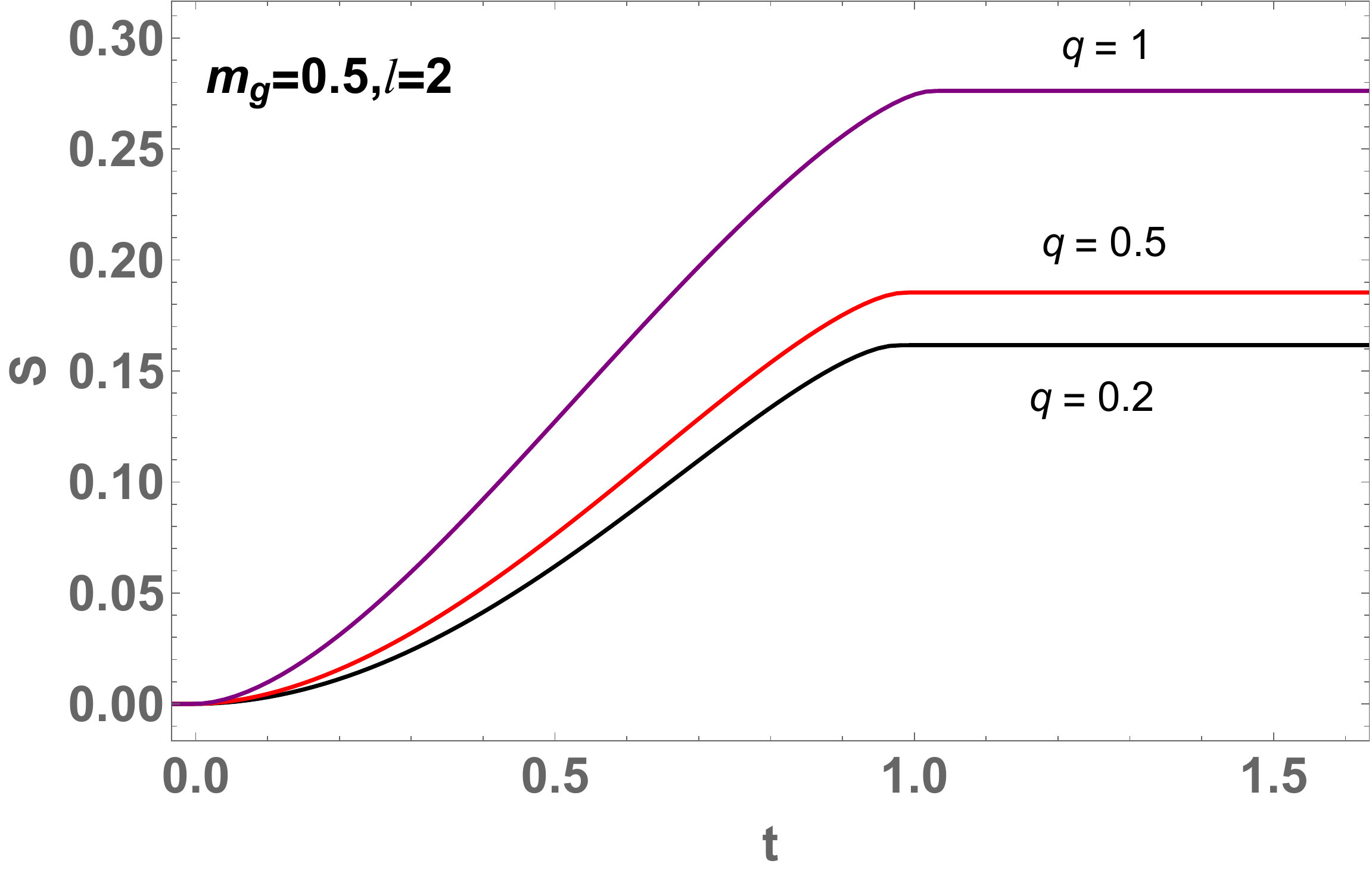}\ \hspace{0.1cm}
  \includegraphics[width=7.8cm]{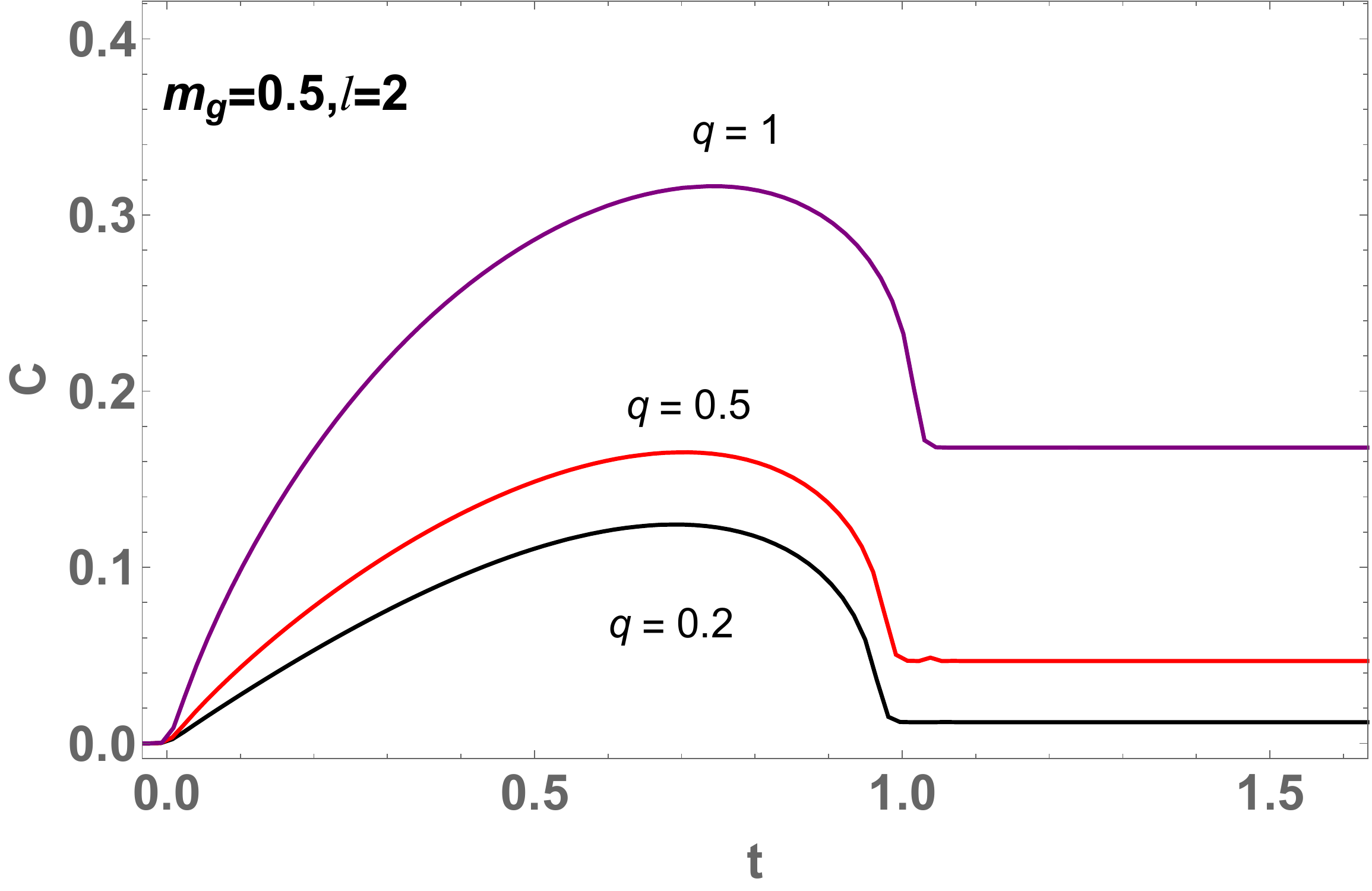}\ \\
	  \caption{Left plot: The evolution of the HEE in massive charged BTZ black hole for different $q$.	
	  Right plot: The evolution of the HC in massive charged BTZ black hole for different $q$. Here we fix $m_{g}=0.5$.}
 \label{fig:massqSEECm05L2}
\end{figure}
\begin{figure}[ht!]
 \centering
  \includegraphics[width=7.8cm]{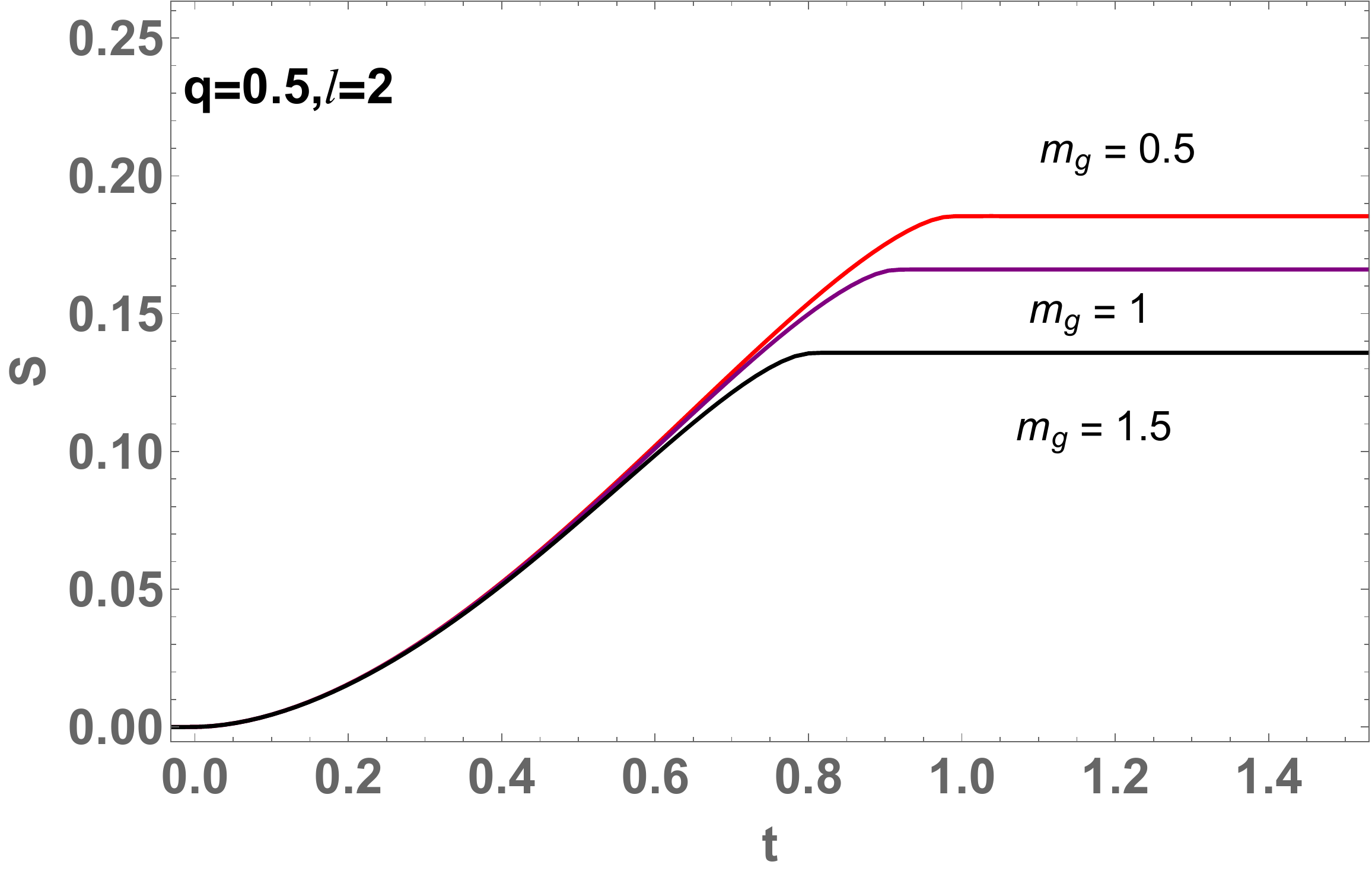}\ \hspace{0.5cm}
  \includegraphics[width=7.8cm]{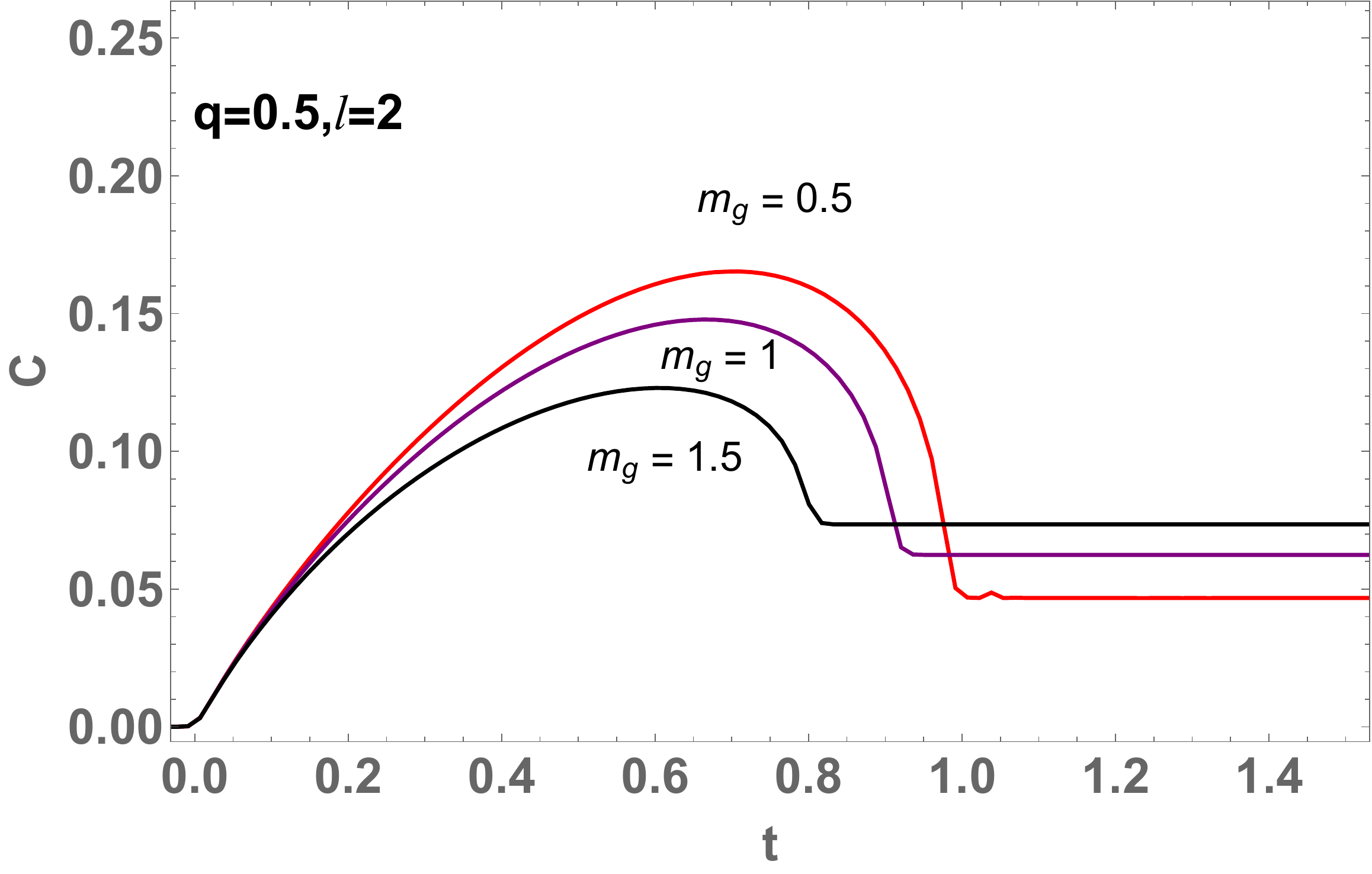}\ \\
	  \caption{Left plot: The evolution of the HEE in massive charged BTZ black hole for different $m_g$.	
	  Right plot: The evolution of the HC in massive charged BTZ black hole for different $m_g$. Here we fix $q=0.5$.}
 \label{fig:massqSEECq05L2}
\end{figure}
\begin{figure}[ht!]
 \centering
  \includegraphics[width=7.8cm]{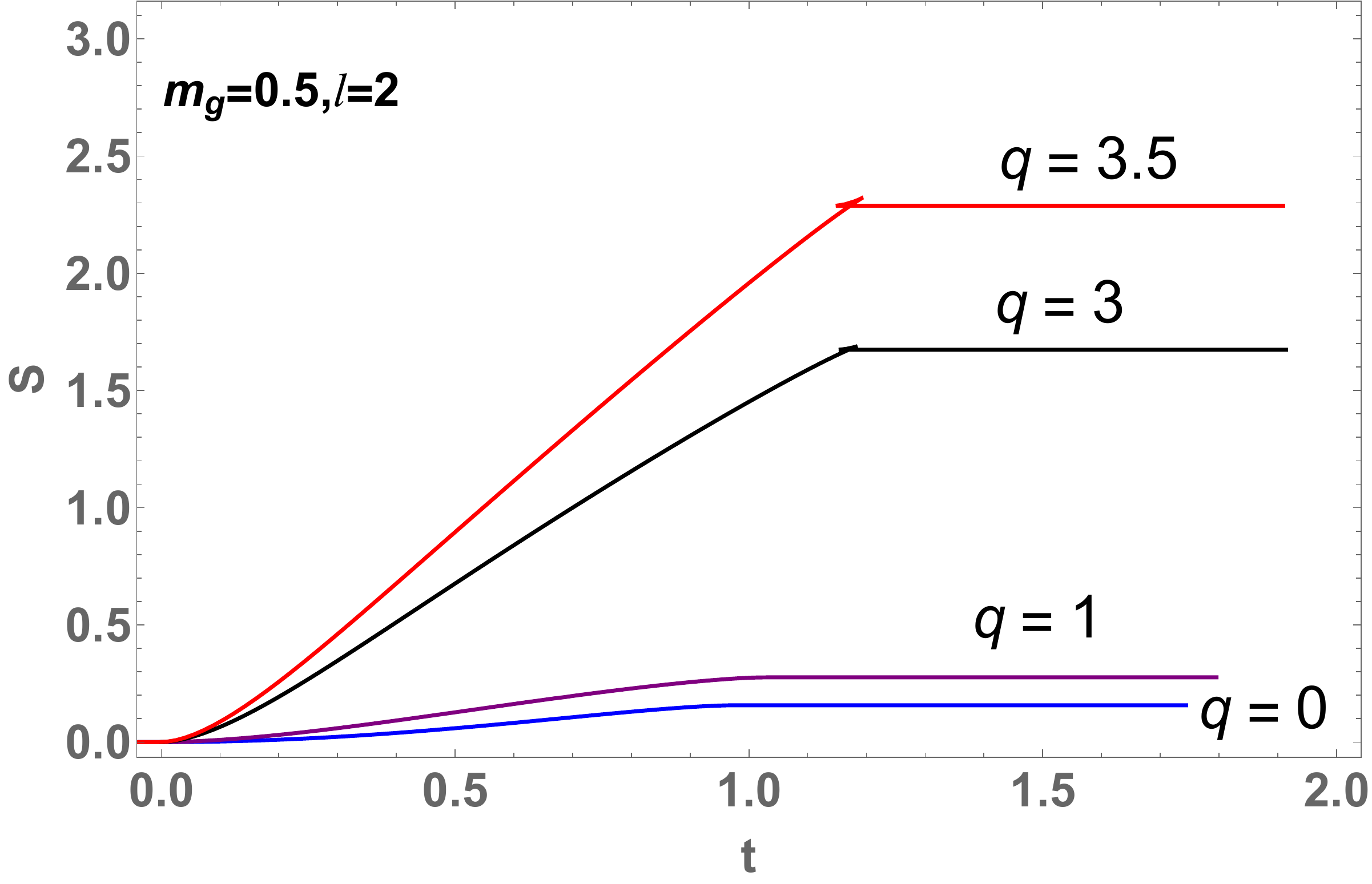}\ \hspace{0.5cm}
  \includegraphics[width=7.8cm]{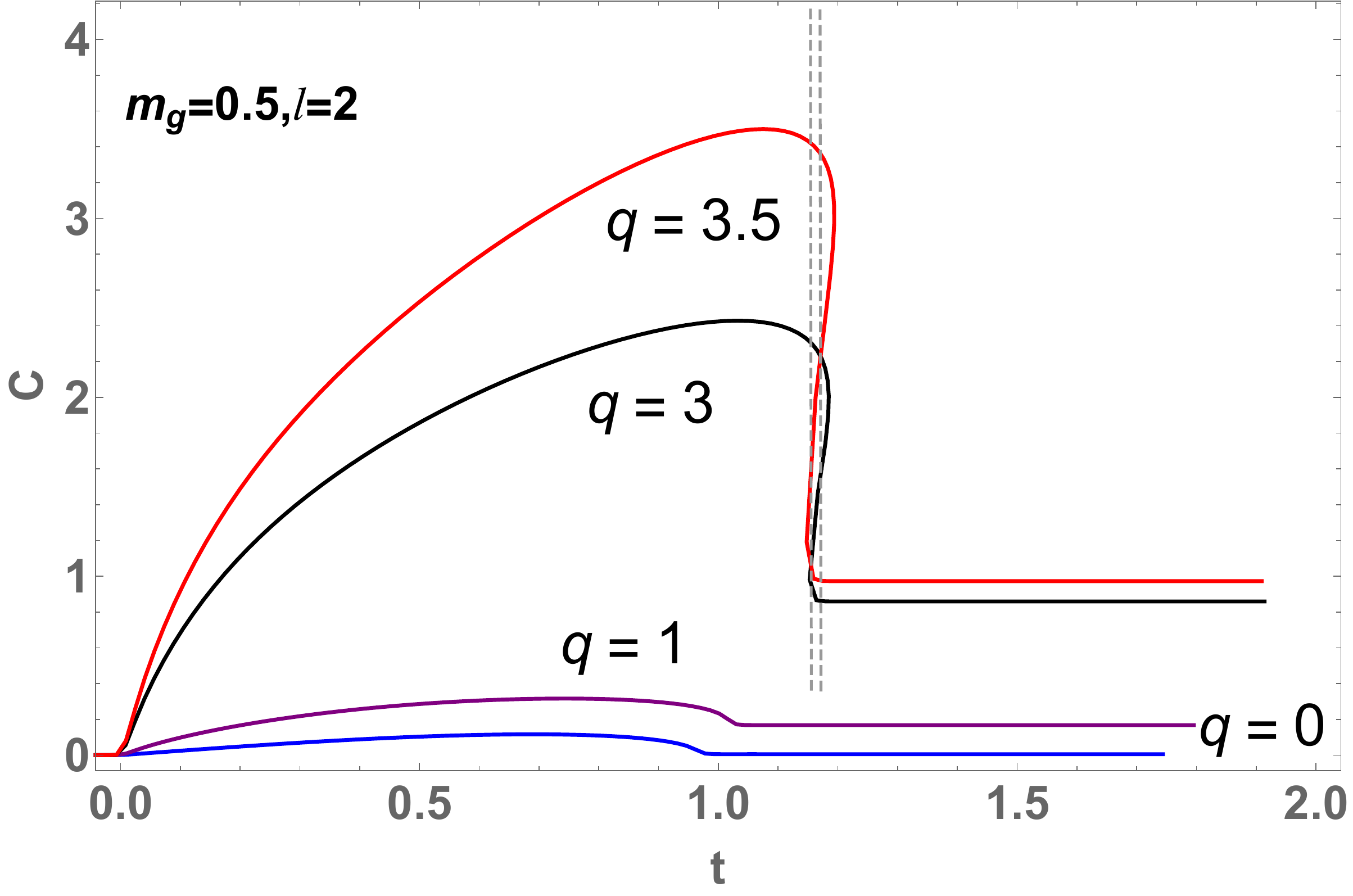}\ \\
	  \caption{The evolution of the HEE (left plot) and the HC (right plot) in massive charged BTZ black hole.
	  Here we have set $m=1$, $m_g=0.5$, $v_0=0.01$ and $l = 2$.}
 \label{fig:massqSEECL2q35q0}
\end{figure}

In this subsection, we briefly discuss the joint effects of the graviton mass and the charge of the black hole on the evolution of the HEE and HC.

The properties of the evolutions of the HEE and HC for small sizes, i.e, $l=2$, are summarized as follows:
\begin{itemize}
	\item The evolutions of the HEE and HC for fixed $m_g$ and different $q$ are exhibited in
	Fig.\ref{fig:massqSEECm05L2}. At the later stage of the evolution, both HEE and HC finally enter into the stable stage.
	By increasing the charge of the black hole, $q$, the final stable values of HEE and HC, and also the maximum value of HC all  increases.
	In addition, for larger $q$, both HEE and HC  take longer times to achieve stability.
	All the properties shown here are closely similar to the case of charged BTZ black hole with the massless graviton, which has been studied in subsection \ref{sub-charge}.
	\item We also present the evolutions of HEE and HC for fixed values of $q$, but different values of $m_g$ in Fig.\ref{fig:massqSEECq05L2}. One could notice that, as $m_g$ increases, the final stable value of the HEE decreases.  The maximum value of the HC also decreases but the stable value of the HC  increases. One could also notice that, the time to achieve the stable region decreases as the parameter $m_g$  increases.
These observations are similar to the case of neutral black hole studied in subsection \ref{sub-neutral}.
	\item One could also observe, specifically from Fig.\ref{fig:massqSEECL2q35q0}, that similar to the case of $m_g=0$, as $q$ increases, the swallow tail of the HEE and also the multi-valuedness of the HC emerge again. In fact, for any specific $m_g$, there exists a critical value of $q$, which beyond that, the swallow tail of the HEE and the multi-valuedness of the HC emerge.
 The main point is that, by just increasing the charge $q$, the swallow tail of the HEE and the multi-valuedness of the HC  emerge, and this behavior is universal, see \cite{Ghodrati:2015rta, Ghodrati:2018hss} and the references therein.
\end{itemize}

\begin{figure}[ht!]
 \centering
  \includegraphics[width=7.9cm]{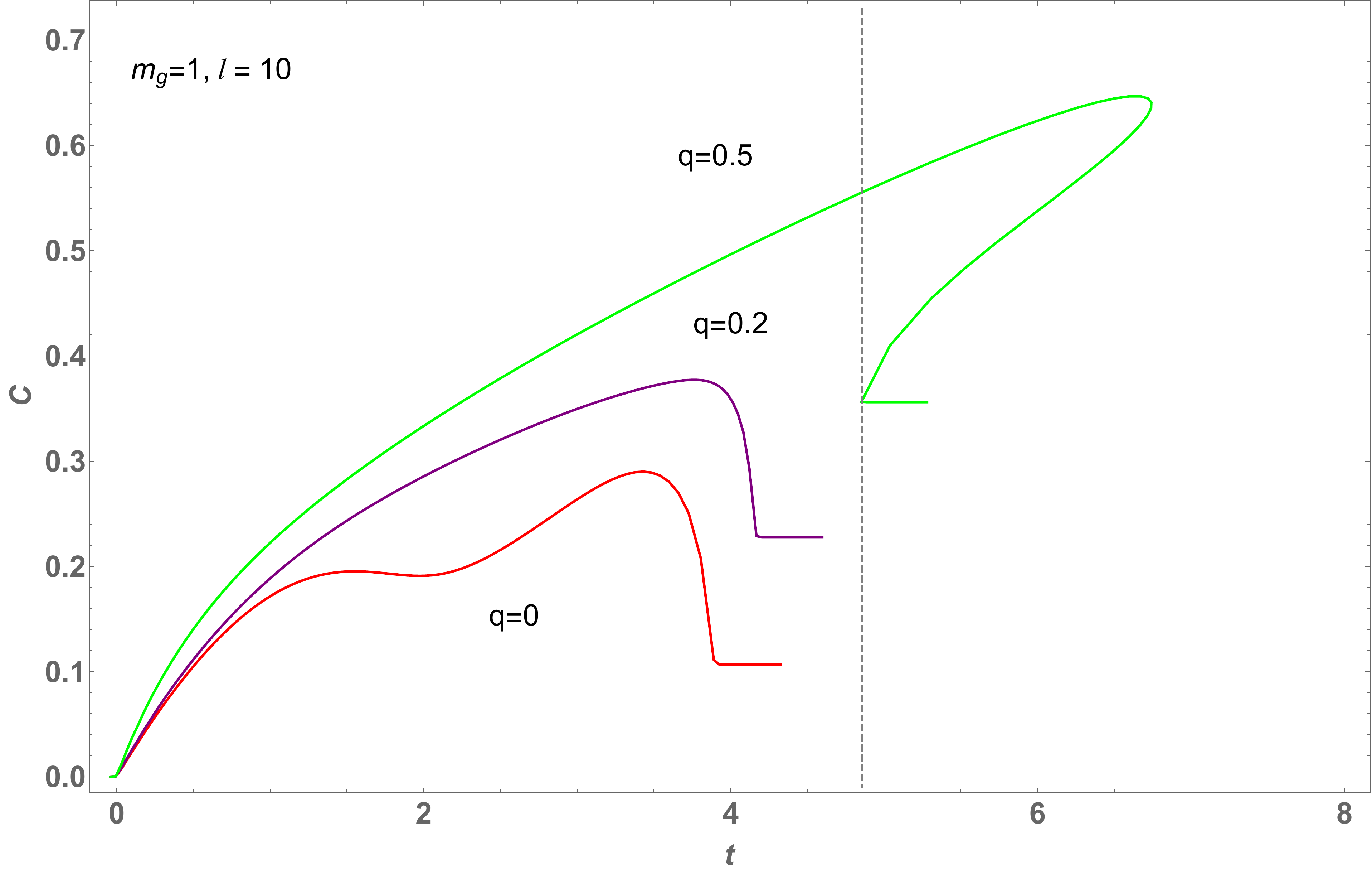}\ \\
  \caption{The evolution of the HC for different $q$, but with fixed $m_{g}=1$. Here we have set $m=1$, $v_0=0.01$ and $l = 10$.}
 \label{fig:ml10ct}
\end{figure}

Then, as for the case of bigger size of strip, for instance $l=10$, of which the result is shown in Fig.\ref{fig:ml10ct}, one could observe that, with large enough charge,  one of the peaks would be smoothed out. This effect is actually introduced by the graviton mass. Also, for the case where the diagrams of HC has one peak, just the same figure of cases with small width will be reproduced.

Another important point is that we check that in all of these examples, the Lloyd's bound conjecture \cite{Brown:2015bva, Brown:2015lvg}, stating
\begin{gather}
\frac{d} {dt} C(t) \le \frac{2}{\pi} E,
\end{gather}
is satisfied. Note that $E$ is the average energy of the state at any time $t$.

One more important point that we have observed by comparing our various plots, is that at early times, and in the first stage of the evolution, the behavior of HC is almost the same for both small and big size of the strip $l$. This point indicates that the growth of complexity is due to the local operator excitations, even when we have dissipations in the system.

Moreover, similar as emphasized in \cite{Chen:2018mcc},we also observed here in our theory and even for the case of massive black hole with massive gravitons, was that both HEE and HC would keep constant after approximately the time $t \gtrsim l/2$, which could be explained using the behavior of thermalization of local states.
As shown in \cite{Cardy:2014rqa}, this is actually due to the fact that after $t \gtrsim l/2$ , the density matrix of subsystem will approach the thermal density matrix exponentially and the correction to the thermal state will just be suppressed as $e^{-4 \pi \Delta_{\text{min} }  (t-l/2)/\beta f(m_g)  }$. Note that here, for our case we should add a function of $m_g$ (or dissipation in the dual field theory) which as we showed affect the final stable value and also the exponential drops. Also, $\beta$ is the inverse temperature and $\Delta_{\text{min}}$ is the dimension of the smallest operator with a non-zero expectation value at the early stage.

Note that these results could have applications in studying the thermalization process in real systems of quark-Gluon and lattice QCD as the effects from the massive graviton in the bulk could be considered as the effects from lattice in the dual field theory.

\section{Conclusion and discussion}\label{sec-IV}

In this paper, we study the evolutions of the holographic entanglement entropy (HEE) and holographic complexity (HC) in the background of massive charged BTZ black holes where the dual CFT goes under a global quench.
We separately explore the effects of the mass of graviton in this theory and the charge of black hole in this solution. After that, we study the joint effects from both of them. We separated the results into two categories of small size and big size of the system.
The qualitative picture we found is summarized as follows:
\begin{itemize}
	\item
	Both the charge of the black hole $q$ and the width of the strip $l$, together  determine whether the evolution of HEE and HC would be a continuous function or not. For small $q$ and $l$, the evolution of the HEE and the HC is always  continuous. We saw that, the HEE climbs up at the first stage of the evolution and then finally it arrives to a stable final region. However, for the case of HC, we saw that it grows until it arrives at a maximum point, and then after that it quickly drops to reach to a stable final stage. When $q$ or $l$ is tuned larger, the discontinuity emerges. We also observed a swallow tail in the evolution of HEE and also noticed that HC is a multi-valued function. These features are special for the charged case and they could not been observed in the neutral AdS$_3$ backgrounds \cite{Chen:2018mcc}.
	\item
	The mass of the graviton plays a crucial role in the evolution of HEE as in the boundary CFT, it corresponds to the dissipations in the system. Its effect is to speed up reaching up to the stability during the evolution of HEE. However, it makes the final stable value lower.
	
	As for the HC, the mass of the graviton also speeds up reaching to the stability during the evolution of the system and it reduces the maximum point, but it raises the final stable value, which is different from the behavior of HEE.  Note that this effect has been observed for the holographic thermalization where the inhomogeneity of the boundary field theory which has been introduced in the bulk by $m_g$ could render the thermalization faster. This  also implies a significant role in defining thermodynamic laws for complexity as well \cite{Bernamonti:2019zyy, Brown:2017jil}.
	
	A novel phenomena that we have observed in the behavior of HC, in charged massive BTZ theory, was that for the systems with large widths, the graviton mass can introduce two peaks in the evolution of HC.  For bigger enough $m_g$, the second peak then evolves to the stable value of HC. Moreover, the charge of the black hole could smooth one of the peaks. For large enough charges, however, the evolution of HC is recovered to usual behavior.
	
	\item
	The emergence of the discontinuity in the HEE and the HC is universal when we tune $q$ or $l$ larger. However, by increasing the graviton mass $m_g$, we could not observe any emergence of the discontinuity, when the charge $q$ and $l$ are small.
\end{itemize}

Moreover, we investigated the evolution of HEE and HC growths for big widths at the early stages, and we found that the growth rates are almost linear. Both larger graviton mass and charge  correspond to higher growth rates for the evolution of HEE and HC.

In addition to the HEE and HC studied here, there are more information related quantities which could be implemented in various setups, such as quenches, and specifically using models with a mass term and therefore dissipations, like the work here, in order to probe various phase transitions in close to real world systems. As we have mentioned in the introduction section,  these quantities  include EoP, CoP and logarithmic negativity, etc.,
each of which  based on their specific characteristics, such as how much they are sensitive to classical or quantum correlations among mixed states in distinct parts of the system, could depict a different, or similar pictures of the phase transitions and evolutions of the system.
EoP  captures both classical and quantum correlations, it is a great quantity to use for probing the phase transitions completely. It has been holographically generalized in \cite{Yang:2018gfq,Liu:2019qje}. Specially when the case is massive, one could check further how it  affects both classical and quantum correlations inside the entanglement wedge \cite{Umemoto:2019jlz, Ghodrati:2019hnn} and then later for the dynamical and quenched systems.
 CoP could also be very sensitive to the dynamics of the system, and may produce the phase structures of specially quantum mixed states and those under time evolutions or quenches.
 Logarithmic negativity  is another quantum information quantity which can be used in detecting phase transitions of different physical systems.  Since this measure is blind to the classical correlations and only captures the ``quantum correlations" with the nature of ``entanglement". Therefore, one would expect the outcomes using this quantity is different from those that come from EoP and CoP.

Other quantum information measures which could be used would be mutual information \cite{Hayden:2011ag,Allais:2011ys,Liu:2019npm}, the R\'enyi entropy \cite{Headrick:2010zt,Belin:2013dva,Dong:2016fnf}, Relative Renyi entropy \cite{Bao:2019aol}, multipartite EoP \cite{Umemoto:2018jpc}, etc.. It would be interesting to study the evolution of all these quantities under the thermal quench using similar methods. By comparing the results from each of these quantum information quantities, many interesting results both about the particular system under study and also the characteristics of the implemented quantum information quantity could be derived.

Recently, in \cite{Caputa:2019avh}, from the other side of story, the problem of quench has been investigated. In that work, however, the quench is a double local, i.e, it is created locally and instantaneously, in a joining or splitting form, and at two different points in the boundary CFT. The authors found that the difference between the double local quench and the sum of two local quenches is negative for various quantities such as energy stress tensor and entanglement entropy which could be interpreted as the gravitational force in the dual gravity theory. This study could be repeated for the complexity and complexity of purification as well.  Specifically, the change in the strength  of this force when the graviton is massive could be calculated.

Other  generalizations of this study, such as  the behavior of HEE and HC  in higher dimensions, higher derivative gravities, composite systems, various quench speeds, or using CA instead of CV would be possible future directions. Comparing the results with other probes such as Wilson loops or two-point correlation functions during quench would also be of interest.

Our work on some of these subjects is under progress.

\begin{acknowledgments}
We appreciate Cheng-Yong Zhang for helpful discussions. This work is supported by the Natural Science
Foundation of China under Grants No. 11705161, 11775036, 11847313, and Natural Science Foundation of Jiangsu Province under Grant No.BK20170481.
\end{acknowledgments}

\end{document}